\documentclass[journal,comsoc]{IEEEtran}

\usepackage[T1]{fontenc}% optional T1 font encoding
\usepackage[utf8]{inputenc}
\usepackage[margin=0.75in]{geometry}
\usepackage{graphicx}
\usepackage{placeins}
\usepackage{float}
\usepackage{bbm}
\usepackage{subfigure}
\usepackage{amssymb}
\usepackage{amsmath}
\usepackage{amsthm}
\usepackage{mwe}
\usepackage[justification=centering]{caption}
\usepackage[font=small]{caption}
\newcommand{\subparagraph}{}
\usepackage[explicit]{titlesec}

\makeatletter
\def\thickhline{%
	\noalign{\ifnum0=`}\fi\hrule \@height \thickarrayrulewidth \futurelet
	\reserved@a\@xthickhline}
\def\@xthickhline{\ifx\reserved@a\thickhline
	\vskip\doublerulesep
	\vskip-\thickarrayrulewidth
	\fi
	\ifnum0=`{\fi}}
\makeatother

\newlength{\thickarrayrulewidth}
\DeclareMathOperator{\E}{\mathbb{E}}
\DeclareMathOperator{\Var}{Var}
\DeclareMathOperator*{\argmax}{argmax}

\title{Deep Reinforcement Learning Control for Radar Detection and Tracking in Congested Spectral Environments}

\newtheorem{definition}{Definition}

\author{Charles~E.~Thornton,~\IEEEmembership{Student Member,~IEEE,}
        Mark~A.~Kozy,~R. Michael Buehrer,~\IEEEmembership{Fellow,~IEEE,}
        Anthony~F.~Martone,~\IEEEmembership{Senior~Member,~IEEE},~and~Kelly~D.~Sherbondy,~\IEEEmembership{Member,~IEEE}% <-this % stops a space
\thanks{C.E. Thornton, M.A. Kozy, and R.M. Buehrer are with Wireless@VT, Department of ECE, Blacksburg, VA, USA. Emails $\{$cthorn14, buehrer$\}$@vt.edu. A.F. Martone and K.D Sherbondy are with the US Army Research Laboratory, Adelphi MD, USA. The support of the US Army Research Office (ARO) is gratefully acknowledged.}}

\begin{document}
% make the title area
\maketitle
\begin{abstract}
This work addresses dynamic non-cooperative coexistence between a cognitive pulsed radar and nearby communications systems by applying nonlinear value function approximation via deep reinforcement learning (Deep RL) to develop a policy for optimal radar performance. The radar learns to vary the bandwidth and center frequency of its linear frequency modulated (LFM) waveforms to mitigate interference with other systems for improved target detection performance while also sufficiently utilizing available frequency bands to achieve a fine range resolution. We demonstrate that this approach, based on the Deep $Q$-Learning (DQL) algorithm, enhances several radar performance metrics more effectively than policy iteration or sense-and-avoid (SAA) approaches in several realistic coexistence environments. The DQL-based approach is also extended to incorporate Double $Q$-learning and a recurrent neural network to form a Double Deep Recurrent $Q$-Network (DDRQN), which yields favorable performance and stability compared to DQL and policy iteration. The practicality of the proposed scheme is demonstrated through experiments performed on a software defined radar (SDRadar) prototype system. Experimental results indicate that the proposed Deep RL approach significantly improves radar detection performance in congested spectral environments compared to policy iteration and SAA. 
\end{abstract}
\begin{IEEEkeywords}
Deep reinforcement learning, cognitive radar, Markov decision process, spectrum sharing.
\end{IEEEkeywords}

%%Charlie Note: Literature Seach Generally finished, might need to clean this section up a bit later
\section{Introduction} 
With the mass deployment of fifth-generation (5G) cellular technology, wireless traffic is heavier and more concentrated than ever before, and all signs are pointed towards continued growth \cite{factsheet}. In tandem, critical applications of radar systems for both government and commercial use have also grown drastically in recent years. Radar and communication systems each require access to significant portions of the sub 6GHz frequency bands, which is planned to be the primary frequency range for 5G systems \cite{release}. Further, short-range radar applications also demand spectrum access in the millimeter-wave bands also proposed for 5G use. Intelligent spectrum access will thus be a significant design consideration for radar systems of the future. Since frequency bands above 1GHz have traditionally been statically allocated for radar use, the vast majority of current radar systems utilize large portions of valuable spectrum in a fixed manner, leading to potential difficulties as spectrum sharing policies are adopted \cite{technical}.

Efficient spectrum sharing from a radar perspective demands robust systems that can make optimal decisions within a heterogeneous EM environment \cite{noncoop}. \textit{Cognitive radar}, a diverse concept first introduced by Haykin in \cite{haykin}, refers to radar systems which utilize feedback information between the transmitter and receiver to optimize parameters of interest. For example, a radar may aim to maximize its own detection performance, or to minimize mutual interference with other spectrum users depending on statistical information about the interference channel \cite{demystified}. Cognitive radar has the ability to meet strict spectral constraints through a combination of adaptive, predictive, and traditional radar processing or waveform optimization techniques \cite{ontheroad}. While fully-aware radar remains an elusive goal for systems to evolve towards, the application of data-driven approaches to next generation radar systems can serve to greatly improve performance.

The applications of cognitive radar are vast, and modern research can typically be broken into two main thrusts. The former focuses on physical layer performance improvement and includes techniques such as transmit waveform optimization \cite{waveformopt}, cognitive beamforming \cite{cogbeamforming}, and resource management \cite{resource}. The latter focuses on spectrum sharing capabilities, and has employed game-theoretic approaches \cite{game}, spectrally contained waveform design \cite{cognotch}, and rapid-reaction Sense-and-avoid (SAA) techniques \cite{SDradar} to mitigate interference. However, the massive amount of received data processed by radars, as well as the ease of utilizing Channel State Information (CSI) due to co-location of the transmitter and receiver, makes the radar control problem a natural candidate for reinforcement learning (RL) algorithms \cite{bell}.

This work proposes a deep reinforcement learning (Deep RL) approach, specifically value function approximation using a Deep $Q$-Network (DQN), to improve the spectrum sharing capabilities of cognitive radar beyond existing approaches. This research extends the approach described in \cite{ersin}, which models a cognitive radar's waveform selection as a Markov Decision Process (MDP), which is a mathematical representation of a sequential decision process \cite{puterman}. An optimal policy is found via a dynamic programming RL algorithm called \textit{policy iteration}, and is shown to mitigate mutual interference between radar and a single communication system. 

Unfortunately, the high-dimensionality of the policy iteration approach limits the number of states the radar can consider, which leads to a limited model of the environment. Similarly, limitations on the number of actions the radar can take will limit the agent's ability to utilize available spectrum efficiently. The Deep RL approach proposed here presents a computationally feasible scheme to deal with these issues. Additionally, the Deep RL techniques presented here allow for real-time updates to learned behavior, so that the radar can continue learning in the case of non-stationary coexistence environments.

Nonlinear function approximation using deep learning models can enable RL to scale to decision-making problems that were previously intractable \cite{deeprlsurvey}. In Deep RL, the computation time per update does not grow with the number of training states, as it does with dynamic programming approaches such as policy iteration. Another previously shown limitation of the policy iteration approach is poor generalization performance in the presence of previously unknown interference states. In the model of \cite{ersin}, in the presence of a previously unobserved state, the radar takes a default action as the state transition model is unspecified for that case. Alternatively, Deep RL allows the radar to take a more informed action based on an estimated function value. 

However, in applying Deep RL to cognitive radar, some domain-specific challenges must be addressed. Radar systems must extract specific target information which is heavily dependent on the operating frequency in addition to target characteristics which may not be known \textit{a priori}. Further, radar is particularly susceptible to interference due to the two-way propagation losses. Many previous works in cognitive radar address a specific setting or application. However, cognitive radar aims to provide reliable performance across a range of settings. Thus, the need for a general Deep RL framework which can address a variety of detection and tracking applications is addressed in our approach.

Another particular challenge is developing a practical scheme for real-time systems. While Deep RL is known to be very capable in synthetic environments, such as video games, the application is not as straightforward when the scheme must be synergistic with other processes, such as signal processing tasks performed by radars. Through simulation and experimentation, we demonstrate that our scheme is realizable, and also highlight key design considerations, such as the limitation of intra-CPI waveform adaptions and the impact of non-stationary environments.

\subsection{Contributions of This Work}
This paper greatly expands on the preliminary work of \cite{kozy}, which presents a basic version of the DQN approach used here and its simulated performance in the presence of simple interference scenarios. In this work we make the following key contributions:
\begin{itemize}	
	\item We provide a full description of the Deep RL model, a demonstration the of the cognitive radar's performance in complex and realistic spectral coexistence scenarios, and a detailed comparison with both the MDP policy iteration approach described in \cite{ersin} and a naive SAA scheme to extensively demonstrate the benefit of Deep RL in cognitive radar coexistence scenarios.
	
	\item We present an extension to the Deep RL approach which utilizes double $Q$-learning and a recurrent LSTM architecture to stabilize the learning process and learn extended temporal correlations when the interference violates the Markov property. This architecture is known as a \emph{Double Deep Recurrent $Q$-Network} (DDRQN).
	
	\item We demonstrate how Deep RL can be practically applied to improve radar detection performance over other cognitive techniques through experiments performed a hardware implementation of prototype radar system.
	
	\item We note that the Deep RL approach exhibits several improvements over policy iteration. First, explicit transition and reward models do not need to be observed in advance. This allows for faster convergence or consideration of more channel state information. Additionally, the radar generalizes more effectively to new environments, as the radar can perform online learning and update its priorities while transmitting in real-time. Finally, the radar develops more efficient action patterns that result in less distortion in the range-Doppler processed data than other cognitive techniques.	
\end{itemize}

\subsection{Organization}
This paper is structured as follows. In Section II, previous work on cognitive radar is surveyed. In Section III, the MDP formulation, Deep RL architecture, and modeling considerations are described. Section IV presents simulation results and a comparison of performance metrics with other schemes. Section V discusses a hardware implementation of the DQN approach on a Software Defined Radar (SDRadar) platform and presents experimental results. Section VI provides concluding remarks.

\section{Related Work}
Cognitive radar research encompasses both performance-oriented approaches and techniques which enable spectrum sharing. Research focused on performance enhancement includes physical layer performance improvements through transmit waveform optimization via a feedback loop \cite{waveformopt}. This allows the transmit waveform can be used to exploit environmental features for more accurate target estimation. For example, a general framework for Bayesian multi-level target tracking approach is introduced in \cite{framework}. Radar Resource Management (RRM) is another major research topic for performance-driven cognitive radar systems \cite{resource}. RRM is used to create priority scheduling for internal processes that compete for limited physical resources. An additional cognitive technique that can be used to increase detection performance is classification of targets from inverse synthetic aperture radar (ISAR) images \cite{isar}, in which information is extracted from two-dimensional high resolution images to discriminate between various targets. 

Within spectrum sharing research for cognitive radar, the literature can again be split into coexistence and cooperation approaches \cite{radarcomms}. Coexistence schemes monitor the spectrum and adjust system behavior to mitigate mutual interference between users, while cooperative schemes employ co-design to jointly optimize the behavior of multiple users. Some cooperative schemes employ a jointly optimized radar-communications node \cite{joint}, while others attempt to treat the spectral scenario as a cooperative game \cite{game}, which requires users to make trade-offs for joint optimization. However, cooperative schemes are often costly to implement and require extensive planning.

The quintessential coexistence approach is Dynamic Spectrum Access (DSA), in which users are assigned primary user (PU) or secondary user (SU) status, and the SU is able to access the spectrum at time increments when the PU is inactive \cite{dsa}. In \cite{stinco}, the authors developed a compressed sensing scheme to reduce the load of spectrum sensing on digital signal processors and used statistical techniques to estimate the behavior of a Primary User (PU) to guide radar transmissions. In contrast to scheduling user transmissions, SAA methods, which adapt the time and frequency locations of radar pulses, have been utilized to reactively avoid interference \cite{SDradar}, \cite{noncoop}. However, SAA assumes that the interference behavior is stationary between time steps, and can be ineffective in scenarios when the channel's coherence time is very short. Another coexistence technique is the introduction of spectral notches in the transmit waveform to attenuate interference \cite{nulls}. However, the waveforms are bandwidth-limited and come at the cost of increased range side-lobes in the coherently processed data \cite{cognotch}. 

To develop more intelligent radar systems, many cognitive radar works have applied adaptive techniques based on the Perception Action Cycle (PAC) \cite{ontheroad}, \cite{SDradar}, which is considered to be a fundamental aspect of animal cognition. The PAC involves a circular flow of information to guide an agent towards a specific goal \cite{pac}. This framework can be used to model the operation of a cognitive radar, which must obtain some information about its environment, act in a logical way, and wait for feedback, which is then fed back to the transmitter. However, these systems must limit processing time to meet strict timing requirements. Thus, statistical and machine learning approaches are warranted for their ability to process data either sequentially or in batches depending on the application.

Previous machine learning applications for cognitive radar systems have focused on the optimization of high-level target tracking systems \cite{tracknn}, \cite{multitrack}, optimization of the target detection threshold \cite{detection}, and network association for radar and communication co-design \cite{wang}. However, the use of deep learning to control real-time waveform selection remains relatively unexplored in the literature, where the numerous recent advancements in RL can be leveraged for joint consideration of spectral efficiency and radar performance.

It has been demonstrated that problems accurately modeled as an MDP can be efficiently solved by architectures such as the Deep-Q Network (DQN) or Recurrent Convolutional Neural Networks (RCNNs) \cite{deeprlsurvey}. The DQN architecture in particular was shown to achieve human level control in an MDP-modeled environment by using a form of $Q$-learning, and stochastic gradient descent (SGD) to update the weights of multiple hidden layers in \cite{natureDQN}. However, performance from this approach is highly dependent on the quality of the radar's assessment of the environment. In real-world applications, the environmental model may result in partial or incomplete observations, which present further challenges. Recently, extensions to the DQN have been proposed to aid in learning sequential data, eliminate estimation bias, and prioritize important experiences with encouraging empirical performance \cite{drql}, \cite{ddql}. This work investigates the necessary considerations when implementing these ideas in a novel cognitive radar prototype system.

\section{System Model and Deep Reinforcement Learning Approach}
This cognitive radar system assumes a model of the radar environment as a MDP, and thus dependence between adjacent time steps due to the Markov property. However, we also test the approach in realistic scenarios, so even if this assumption doesn't hold, we believe the approach is effective and the problem can be considered \textit{nearly Markovian}. Additionally, a recurrent architecture is used in the proposed DDRQN to learn interference with extended temporal correlation. We now proceed to a description of the radar environment and the MDP formulation, followed by details of the DQN architectures used to perform Deep RL. 

\subsection{Coexistence Environment and MDP Formulation}
\begin{figure}[t]
	\centering
	\includegraphics[scale=0.55]{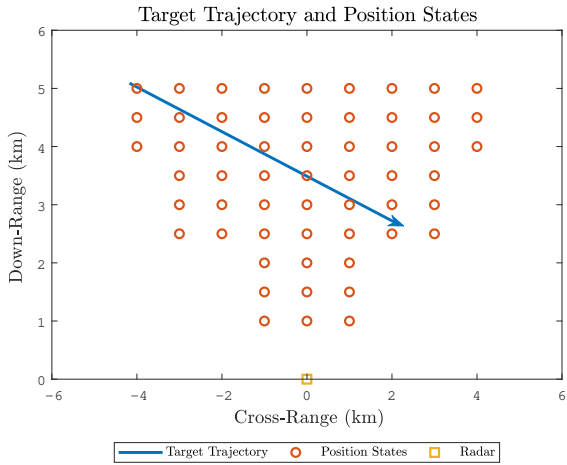}
	\caption{A Down-Range example of the radar scene and target trajectory.}
	\label{fig:traj}
\end{figure}

Consider a radar environment which consists of a single point target, which is being tracked by the radar over an episode consisting of many time steps, where each time step corresponds to a radar pulse. The target has some randomly defined trajectory for each individual episode. The EM environment is a shared channel, split into $N$ equally sized sub-channels. The environment contains a communications system which is capable of operating in one or more sub-bands concurrently and may cause interference to the radar. The radar system is monostatic and stationary, and operates using a Linear Frequency Modulated (LFM) chirp waveform with an appropriate time bandwidth product\footnote{The time bandwidth product is given by $TBP = T(f_{hi}-f_{lo})$ where $T$ is the pulse length and $f_{hi}-f_{lo}$ is the frequency swept by the chirp waveform.}. Losses due to atmospheric effects or clutter are assumed to be negligible. An example of the radar scene is shown in Figure \ref{fig:traj}. The orange circles represent possible target positions, and the arrow represents the trajectory of the target. The target position is a continuous variable, but we use quantized values to maintain tractability.     

A MDP is a well-studied mathematical framework for sequential decision making under uncertainty \cite{puterman}. The model is specified by the tuple $(\mathcal{S}, \mathcal{A}, T, R, \gamma)$. The framework consists of states, actions, transition probabilities, a reward function, and the discount factor. The state $s \in \mathcal{S}$ is a unique collection of important information in the problem being modeled. Here, the state is a vector which includes the target's position $\mathbf{x} \in \mathcal{X} = \{\mathbf{x}_1, ..., \mathbf{x}_P\} $, the target's velocity $ \mathbf{v} \in \mathcal{V} = \{\mathbf{v}_1, ... , \mathbf{v}_V\} $, and the interference due to communication systems $\boldsymbol{\theta} \in \boldsymbol{\Theta} = \{\boldsymbol{\theta}_1, ... , \boldsymbol{\theta}_M \}$, where $P$ is the number of position states, $V$ is the number of velocity states, and $M$ is the number of unique interference states.

Each interference state $\boldsymbol{\theta}_{i} \in \boldsymbol{\Theta}$ is expressed as a length $N$ vector of binary values in which a $1$ corresponds to an occupied sub-channel and $0$ denotes availability. For example, in the case of $N = 5$ sub-bands $\boldsymbol{\theta} = [1,1,1,0,0]$ corresponds to the first three sub-channels being utilized and the remaining sub channels are available for radar use. Given \textit{N} sub-bands, the total number of unique interference states is then $M = 2^N$. The total number of states is thus $N_s = P \times V \times 2^N$, which grows exponentially as the number of sub-bands increases, limiting our discretization of the channel in practice. 

The radar's action space $\mathcal{A}$ consists of transmissions in any grouping of contiguous sub-bands. Similar to the interference state $\boldsymbol{\theta}$, each radar action $\mathbf{a}_{i} \in \mathcal{A} = \{\mathbf{a}_{1}, ..., \mathbf{a}_{N_{a}} \}$ is formulated as a length $N$ vector of binary values in which a $1$ corresponds to the presence of the radar's chirp waveform and $0$ denotes no radar activity. Since the radar is limited to occupying contiguous sub-channels, the number of actions\footnote{Since the radar is limiting to transmissions in contiguous sub-bands, the number of possible transmissions will always be a triangular number, yielding the $N(N+1)/2$ result.} is then $N_a = \frac{N(N+1)}{2}$.

The transition probability function $T(s,a,s^{\prime}): \mathcal{S} \times \mathcal{A} \times \mathcal{S} \rightarrow [0,1]$ is the probability of reaching state $s^{\prime}$ by taking action $a$ while in state $s$. The MDP framework assumes the Markov property holds for the environment, and thus only single-step transitions are considered. However, this assumption will be relaxed in the Deep RL approach. The reward function $R(s,a,s^{\prime}): \mathcal{S} \times \mathcal{A} \times \mathcal{S} \rightarrow \mathbb{R}$ specifies a numerical value that determines the agent's preference for taking some action while in a particular state, and is defined by the system planner. The unique combination of the reward and transition functions characterizes the MDP model. 

Finally, the discount factor $\gamma \in [0,1]$ is a weighting term which specifies the radar's preference for near-term ($\gamma \approx 0$) or long-term ($\gamma \approx 1$) rewards. The goal of RL is to maximize the \textit{discounted return}, given by a weighted sum of rewards
\begin{equation}
G_{t} = \sum_{k = 0}^{\infty} \gamma^{k} R_{t+k+1},
\end{equation} 
\noindent where $R_{t+k+1}$ is the reward at time $t+k+1$. Choosing a reward function that corresponds to ideal interaction with the environment is a critical aspect of any RL approach, so care will be taken here. The radar must value utilizing as much available spectrum as possible to obtain a fine range resolution. Additionally, we wish to minimize interference to other systems and improve target detection capabilities, so instances when the radar occupies frequency bands concurrently with the communications system must be discouraged. Previous work has found that rewards based on a weighted combination of SINR and bandwidth serves to increase the target detection capabilities of the radar \cite{ersin}. This type of reward function has the form
\begin{equation}
R_{t} =
\left \{
\begin{array}{ll}
\alpha_{1}(\operatorname{SINR})+\alpha_{2}(\operatorname{Bandwidth}), & \operatorname{SINR} \geq 0 \\
\operatorname{Large \; Penalty}, & \operatorname{SINR} < 0\\
\end{array} 
\right \},
\label{eq:ersin}
\end{equation}
\noindent where $\alpha_{1}$ and $\alpha_{2}$ are parameters. As SINR corresponds to the prominence of the target in range-Doppler processed data and larger bandwidth is associated with finer range resolution, this trade-off is intuitive in terms of radar performance. However, reliable SINR measurements require knowledge of the target's range and assumptions about scattering properties. Further, depending on the values of $\alpha_{1}$ and $\alpha_{2}$, the radar may receive a sufficiently high reward by utilizing the entire channel, which may not be conducive to coexistence scenarios.

To foster both interference avoidance and bandwidth utilization for spectrum sharing, this formulation instead uses a reward function based on \textit{missed opportunities} and \textit{collisions}.
\begin{definition}		
	Let the number of collisions $N_{c}$ correspond to the number of sub-channels utilized by both the radar and communication system\footnote{In practice, the radar can calculate collisions by estimating SINR and denoting a collision as the event $(SINR < \Gamma)$. Note that an explicit value of SINR is not required to calculate the reward, as in (\ref{eq:ersin}).}, $N_{c} = \sum_{i = 1}^{N} \mathbbm{1} \{ \mathbf{a}_{t,i} = \boldsymbol{\theta}_{t,i} \}$, where $\mathbbm{1}\{\cdot\}$ is the binary indicator function and the notation $\mathbf{a}_{t,i}$ corresponds to the $i^{th}$ element of the action taken at time $t$.
\end{definition}

\begin{definition}
	Let the number of missed opportunities $N_{mo}$ correspond to the difference between the largest group of contiguous sub-channels not utilized by other systems \footnote{In practice, $\mathbf{a}^{*}_{t}$ could be calculated using an threshold-based energy detector}, $\mathbf{a}^{*}_{t}$, and the number of those sub-channels the radar actually selects, $N_{mo} = \lVert \mathbf{a}^{*}_{t} \rVert - \sum_{i = 1}^{N} \mathbbm{1} \{ \mathbf{a}^{*}_{t,i} = \boldsymbol{a}_{t,i} \}$.
\end{definition}

The radar's reward function $R_{t} \in [0,1]$ can then be expressed as
\begin{equation}
R_{t} =
\left \{
\begin{array}{ll}
0 & \hspace{0.95cm} N_{c} > 0 \\
\beta_{1}/({\beta_{2}} N_{mo}) & \hspace{0.95cm} N_{c} = 0, N_{mo} > 0 \\
1 & \hspace{0.95cm} N_{c} = 0, N_{mo} = 0 \\
\end{array} 
\right \},
\label{eq:rwd}
\end{equation}
\noindent where $0 \leq \beta_{1} < \beta_{2}$ are parameters which specify the radar's preference for interference avoidance relative to utilization of available bandwidth. Large values of $\beta_{1}/\beta_{2}$ emphasizes interference avoidance, as the radar can achieve close to the maximum reward despite $N_{mo} > 0$. Alternatively small $\beta_{1}/{\beta_{2}}$ encourages the radar to attempt to utilize full available channel whenever possible as missed opportunities drive the reward closer to $0$. The choice of reward parameters is highly application-dependent, as range resolution and target detection goals will dictate the radar's preferences. 

This reward function proves to be effective, since the radar will quickly learn that colliding with radio-frequency interference (RFI) yields the minimum reward. Thus, the radar will learn to avoid actions that regularly result in collisions. To obtain the highest possible reward, the radar must use the greatest number of available contiguous bands, which encourages use of the available spectrum. Further, $N_{c}$ and $N_{mo}$ are easily calculable at each time step and bounded by $N_{c} \leq N$ and $N_{mo} \leq N-1$ respectively, promoting a stable learning process and learnable association between states and actions. 

The goal of a MDP is to learn a mapping between states and a probability distribution over actions, known as policy $\pi$. The optimal policy, which yields the highest expected \textit{discounted} reward is denoted by $\pi^{*}$. We now describe how the optimal policy can be found iteratively using a dynamic programming approach known as policy iteration and more practically approximated via our Deep RL approach.

\subsection{Policy Iteration Approach}
We now briefly describe the policy iteration approach as it was implemented in \cite{ersin} and will be used as a baseline for comparison to the Deep RL approach. This technique involves estimation of $T(s,a,s')$ through repeated experience followed by the well-known policy iteration algorithm. Policy iteration is a dynamic programming technique performed by starting from arbitrary policy $\pi_{o}$ and iteratively performing two steps, \textit{policy evaluation} and \textit{policy improvement}. During policy evaluation, the value function $V^{\pi}(s)$, which describes the utility of being in state $s$ while acting on policy $\pi$ is
\begin{equation}
\begin{split}
V^{\pi}(s) &= \E_{\pi}[G_{t}|S_{t} = s] \\ &= \E_{\pi} \left \{\sum_{k = 0}^{\infty} \gamma^{k} R_{t+k} \bigg| S_{t} = s \right \}, \hspace{0.4cm}  \forall \; s \in \mathcal{S}.
\label{bell}
\end{split}
\end{equation}

Policy improvement then uses $V^{\pi}(s)$ to choose an action $a$ such that the expected utility of the next state $s'$ is maximized. This process creates a new policy, $\pi'$ with an updated value function $V^{\pi'}$ which is guaranteed to be greater than or equal to $V^{\pi}$ by the policy improvement theorem \cite{sutton}. This process then repeats until we obtain a stable solution $V^{*}(s)$ given by
\begin{equation}
\begin{split}
V^{*}(s) & = \max_{a} \E \left[R_{t+1}+\gamma V^{*}(s^{\prime}) \bigg| S_{t} = s, A_{t} = a \right] \\
& = \max_{a} \sum_{s^{\prime}, r} \operatorname{Pr}(s^{\prime},r|s,a)[r + \gamma V^{*}(s^{\prime})],
\end{split}
\end{equation}
from which we take 
\begin{equation}
\pi^{*}(s) = \argmax_{a} \sum_{s^{\prime}, r} \operatorname{Pr}(s^{\prime},r|s,a)[r + \gamma V^{*}(s^{\prime})], 
\end{equation}
to be an optimal policy\footnote{The optimal value function is guaranteed to be unique, but there may be multiple optimal policies.}. In practice, a stopping condition is used to terminate the algorithm once changes become very small.

Policy iteration will converge to $V^{*}(s)$ in finite time \cite{sutton}, but given the large and often sparse, $T(s,a,s^{\prime})$ and $R(s,a,s^{\prime})$ matrices necessary to perform the computations, the approach is not computationally feasible for large state-action spaces. Further, to compute $\pi^{*}$ the radar must already have an accurate model of the transition and reward dynamics built up through repeated experience, which can take many iterations and hinder time-sensitive radar applications, such as target tracking. Finally, once $\pi^{*}$ is found using this approach, the radar's performance is fixed unless the algorithm is run again. For a more practical approach when the environment's dynamics are unknown, we look to function approximation via Deep RL.
\begin{figure*}[t]
	\centering
	\includegraphics[scale=0.55]{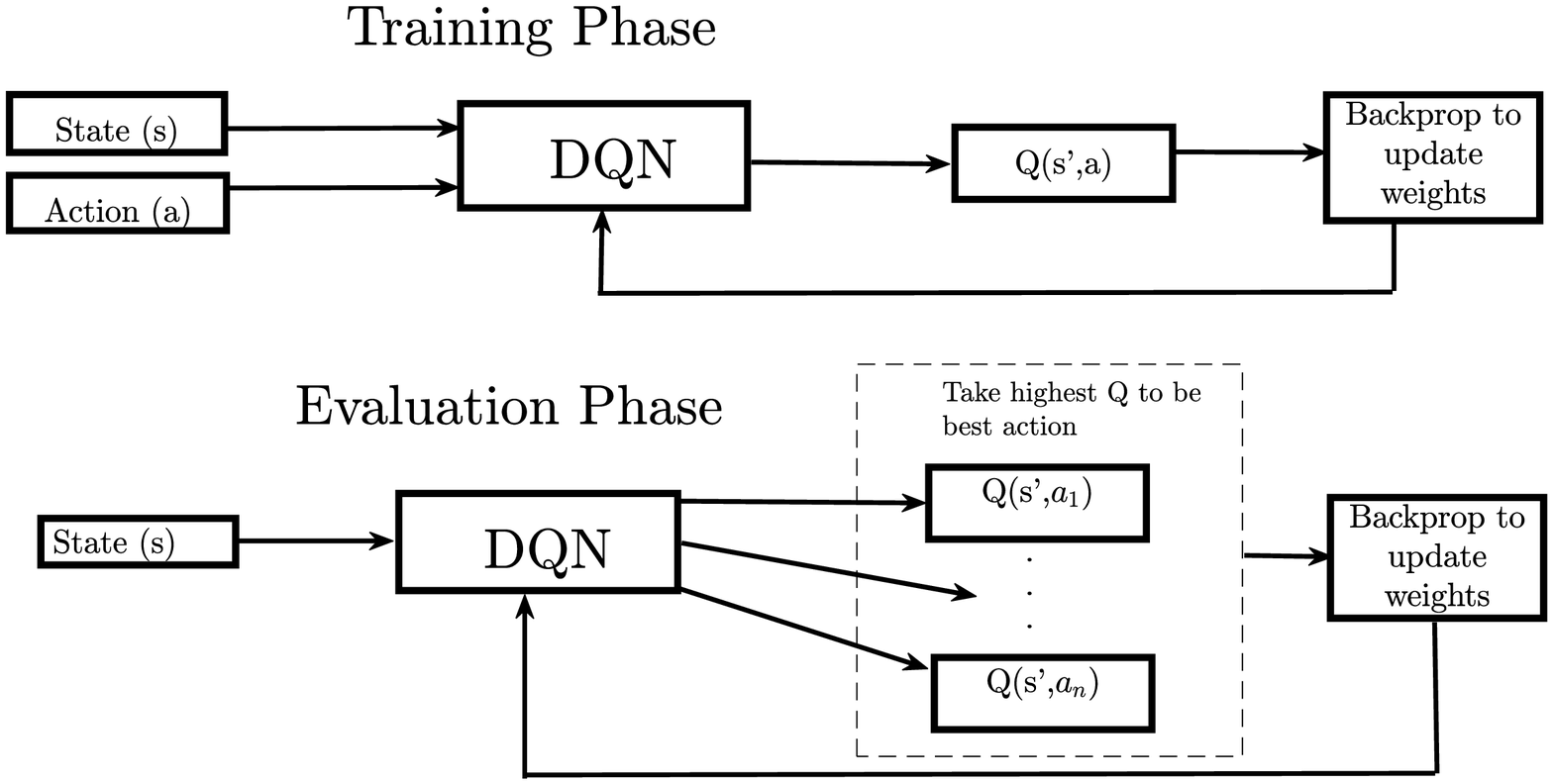}
	\caption{Training (top) and evaluation (bottom) stages of RL agent operation. During training two neural networks are used to estimate the \textit{Q}-value for each state-action pair. During the evaluation phase, an action is selected based on the estimated \textit{Q}-value.}
	\label{fig:dqn}
\end{figure*}

\subsection{Deep RL Approach}
The proposed Deep RL cognitive radar approach is based on Deep $Q$-learning (DQL), an algorithm which attempts to find an approximate solution to a MDP by directly learning the state-action value function $Q(s,a)$, which maps state-action pairs to rewards \cite{sutton}. $Q(s,a)$ is similar in nature to the state value function $V(s)$ above and is expressed by
\begin{equation}
Q^\pi(s,a) = \E_{\pi} \left[ \sum_{k=0}^{\infty} \gamma^{k} R_{t+k+1} \bigg| S_{t} = s, A_{t} = a \right],
\end{equation}
which corresponds to the expected utility of action $a$ in state $s$ while following $\pi$. The optimal $Q$ function, $Q^{*}(s,a)$ is then 
\begin{equation} 
\begin{split}
Q^{*}(s,a) & = \sup_{\pi} Q^{\pi}(s,a) \\
& = \E_{s' \sim \mathcal{S}} \left[r+\gamma \max _{a^{\prime}} Q^{*}(s^{\prime}, a^{\prime}) \big| S_{t} = s, A_{t} = a\right] \\
& = \sum_{s^{\prime}, r} \operatorname{Pr}(s^{\prime},r|s,a) [r + \gamma \max_{a^{\prime}} Q^{*}(s^{\prime},a^{\prime})].
\end{split}
\end{equation}	

DQL approximates $Q^{*}(s,a)$ directly, without examining all policies. This is done using gradient updates via the backpropagation algorithm to update the neural network weights based on the radar's experiences. SGD estimates the gradient of a function and updates the weights, $\mathbf{w}_t$, based upon a single randomly chosen sample $\mathbf{z}_t$. The general form of an SGD update is
\begin{equation}
\mathbf{w}_{t+1} =  \mathbf{w}_t - \alpha_{t} \triangledown_\mathbf{w} L(\mathbf{z}_t,\mathbf{w}_t),
\end{equation}
where $\alpha_{t}$ is the step size or learning rate, $\triangledown_{\mathbf{w}}$ corresponds to the gradient with respect to weight vector $\mathbf{w}$, and $L(\cdot)$ is the loss function \cite{goodfellow}. To improve the accuracy of gradient estimation, we use a method called \textit{experience replay} \cite{natureDQN}. At each time step $t$, the radar experiences a transition
\begin{equation}
\phi_{t} = (s_{t}, a_{t}, r_{t+1}, s_{t+1}),
\end{equation}
\noindent which is then stored in a memory buffer $\mathcal{M}$ with a large, but fixed capacity. A batch of uniformly sampled transitions is used to perform training updates. This method aims to obtain uncorrelated samples which improve the gradient estimate, since strongly correlated updates break the inherent i.i.d assumption of the SGD algorithm. The replay memory also allows the radar to remember rare, but potentially important experiences that occurred many time steps ago.

Value function approximation using neural networks is often unstable due to rapidly fluctuating $Q(s,a)$ estimates. To combat instability, DQNs utilize a \textit{target network} to smooth the learning process \cite{natureDQN}. In this approach, a target network is used to contain the weights that are used to select the best action, denoted by the vector $\mathbf{w}^{\prime}$. The target network's weights are frozen for $T_{target}$ time steps while a separate policy network with weight vector $\mathbf{w}$ responds to every observed state. After the $T_{target}$ steps are complete, the target network's weights are adjusted $\mathbf{w}^{\prime} \leftarrow \mathbf{w}$. This allows the policy network to calculate errors without utilizing its own rapidly fluctuation estimates of the \textit{Q}-values to select waveforms. 

The loss function used to perform learning is based on the standard temporal-difference update. The two networks are said to be the slowly updating target network, with weight vector $\mathbf{w}^{\prime}$, and the policy network which is updated every time step, with weight vector $\mathbf{w}$. At each iteration $i$, we draw $n$ approximately i.i.d. samples from the memory buffer  $\{\phi_{i} \}_{i \in \mathcal{M}}$ and form the target
\begin{equation}
Y_{i}= r_{i+1} +\gamma \max _{a^{\prime} \in \mathcal{A}} Q\left(s_{i+1}, a^{\prime}; \mathbf{w}_{i}^{\prime} \right),
\end{equation}
\noindent where the notation $Q(s,a;\mathbf{w}_{i}^{\prime})$ corresponds to the current $Q$-value estimated by the frozen target network with weight vector $\mathbf{w}_{i}^{\prime}$. The loss function at each step $i$ is then 
\begin{equation}
\label{eq:loss}
\begin{split}
L_{i}(\mathbf{w}_{i}) &= \E_{s,a,r,s^{\prime}} [(Y_{i} - Q(s_{i}, a_{i}; \mathbf{w}_{i}))^{2}]\\
&\approx \frac{1}{n} \sum_{i = 1}^{n} [(Y_{i} - Q(s_{i}, a_{i}; \mathbf{w}_{i}))^{2}],
\end{split}
\end{equation}
\noindent where $Q(s_{i},a_{i};\mathbf{w}_{i})$ corresponds to the policy network's approximated $Q$-value based on the most recent experience and the expectation is approximated using the $n$ samples from memory. SGD updates can then be performed using the gradient of (\ref{eq:loss}). To further analyze the loss function, we can apply bias-variance decomposition 
\begin{multline}
\label{eq:bvar}
\E [L_{i}(\mathbf{w}_{i}) ] = \lVert Q(s_{i},a_{i};\mathbf{w}_{i}) - (r_{i+1} + \\\gamma \E[ \max_{a'} Q(s_{i+1},a',\mathbf{w}^{\prime}_{i})]) \rVert^{2} + \Var [Y],
\end{multline}
where the first term is the mean Bellman squared error (MBSE), a common measure of distance between the estimated and true value functions \cite{sutton}. We further note the variance of the targets $Y$ is dependent on $\mathbf{w}_{i}^{\prime}$ but not $\mathbf{w}_{i}$, allowing us to ignore the variance term. Thus, minimizing (\ref{eq:loss}) is close to finding the minimum MSBE. Without the use of a target network, the variance term in (\ref{eq:bvar}) is dependent on $\mathbf{w}_{i}$ and minimizing (\ref{eq:loss}) can diverge from minimizing the MSBE \cite{understanding}. Thus, both the target network and experience replay are important tools for stability which can be theoretically justified.

The DQN utilized for the experiments in this paper is built in Python using the TensorFlow library \cite{tensorflow}. The network architecture consists of an input layer, three fully-connected hidden layers, and an output layer. While some implementations input several previous states to the network, we input a single state only, as the interference scenarios considered are often stochastic. Rectified linear unit (ReLu) activiation functions are applied at the output of each layer. Specific hyperparameter choices used for both the neural network architecture and radar scene are specified in Section \ref{se:sim}.

Since the DQN consists of many weights which must be learned directly from experience, we utilize an offline training, or exploration period, to develop an initial approximation of $Q(s,a)$. During this phase, the radar selects actions with uniform probability, observes the associated $\phi(s_{t},a_{t},r_{t+1},s_{t+1})$ transition, and updates the memory buffer used to perform gradient updates. In a critical application, the radar would likely simulate taking each action to avoid the poor performance associated with random exploration. The radar the enters an online evaluation phase, where at each step action $A_{t} = \argmax_{a} Q(s,a)$ is taken. Transitions are still stored in the memory buffer and used to update the network's weights, allowing adaptation in non-stationary environments, or continued learning if the experiences gathered in the exploration phase were not sufficient to learn the environment. The training and evaluation phases are depicted in Figure \ref{fig:dqn}.

Deep $Q$-learning presents several important theoretical advantages over the policy iteration approach. Foremost, explicit transition and reward models are not necessary, and radar behavior can be learned based on whatever experience is available. Further, the reduced dimensionality of the problem allows for larger state-action spaces to be considered, increasing the radar's predictive power in interference scenarios with extended temporal dependencies. Finally, the radar is able to continue updating its beliefs during evaluation, which is advantageous in cases of dynamic environments or insufficient training time. However, since transition and reward models are not available, the DQN is heavily dependent on the quality of observed transitions and associated estimates. Since coexistence scenarios are often dynamic and spectrum sensing often yields noisy measurements, we present an extension to the DQN in the next section.

\subsection{Double Deep Recurrent $Q$-Networks}
The DQN model is heavily reliant on the quality of observations and estimates for accurate function approximation. Since real-world observations are often noisy or unreliable, we combine two important extensions to Deep $Q$-Learning in this section for improved performance for the spectrum sharing domain. The first technique is the use of a Double DQN (DDQN), which mitigates estimation bias in the learning process \cite{ddql}. Since the loss function in (\ref{eq:loss}) involves taking a maximum over estimated values, the DQN approach often suffers from overoptimistic $Q$-value estimates due to maximization bias \cite{sutton}. The DDQN approach mitigates overoptimism by using the loss function
\begin{multline}
\label{eq:ddqn}
L_{i}(\mathbf{w}_{i}) = [r_{i+1}+\gamma Q(s_{i+1}, \underset{a^{\prime}}{\operatorname{argmax}}  Q\left(s_{i+1}, a^{\prime}; \mathbf{w}_{i} \right);\mathbf{w}_{i}^{\prime}) \\ - Q \left( s_{i}, a_{i}; \mathbf{w}_{i} \right) ]^{2},
\end{multline}
\noindent which, similar to the DQN, utilizes the policy network weights $\mathbf{w}_{i}$ to select actions, but also uses the frozen target network weights $\mathbf{w}^{\prime}_{i}$ to evaluate the value of the selection. This approach has been shown to significantly mitigate overoptimism associated with the DQN, promoting stability \cite{ddql}.

The second extension is the use of a recurrent neural network (RNN) architecture to process sequences of data such that long term dependencies can be resolved \cite{lstm}. In addition to fully-connected hidden layers described previously, this configuration uses a Long Short-Term Memory (LSTM) layer between the final fully-connected hidden layer and the linear output layer. This architecture known as a Deep Recurrent $Q$-Network (DRQN) \cite{drql}. The DRQN has demonstrated impressive empirical performance in spectrum access tasks with underlying partially observable or noisy state observations, which can be generalized as partially observable MDPs (POMDPs) \cite{yue}. Here, we use the Keras API \cite{keras} to build the fully-connected and LSTM layers that form the DDRQN. Similar to the aforementioned DQN, we utilize three fully-connected hidden layers and ReLU activation functions.
\begin{table*}[t]
	\footnotesize
	\centering
	\caption{Summary of Main Simulation Parameters}
	\label{tab:simParam}
	\begin{tabular}{||l|l||l|l||}
		\hline
		\textbf{Parameter} & \textbf{Value} & \textbf{Parameter} & \textbf{Value} \\ \hline
		Memory Buffer Size & 2000 Transitions & Learning Rate $\alpha$ & 0.001 \\ \hline
		Batch Size & 32 & Target Network Update & Every 250 Steps\\ \hline
		Shared Channel Bandwidth & 100MHz & Sub-Channel Bandwidth & 20MHz \\ \hline
		Target RCS & 0.1 m$^{2}$ & Discount factor $\gamma$ & 0.9 \\ \hline
		Fully-Connected Layer Sizes & (256, 128, 84) Neurons & LSTM Size (DDRQN) &  84 \\ \hline
		Episode Length (DDRQN) &  10 & Reward Parameters $(\beta_{1}, \beta_{2})$ & (5, 6) \\ \hline
		Position States $P$ & 50  & Velocity States $V$ & 10  \\ \hline
		Coherent Processing Interval (CPI) & 1000 Pulses & Pulse Repetition Interval (PRI) & 0.41 ms \\ \hline
	\end{tabular}
\end{table*}

Interference to radar by communications systems will often have non-Markovian structure. Also, spectrum sensing techniques may result in noisy or unreliable state measurements. Thus, there is a need for both stability and resolution of long-term dependencies in the cognitive radar learning process. We propose utilizing the DDQN loss function given by (\ref{eq:ddqn}) in tandem with the DRQN network architecture to form a Double DRQN (DDRQN). Despite being a simple extension of these two techniques, the DDRQN demonstrates superior performance and stability upon convergence in the cognitive radar setting, as shown in the next section. 

\section{Simulation Results} \label{se:sim}
\begin{figure*}[t]
	\centering
	\includegraphics[scale=0.55]{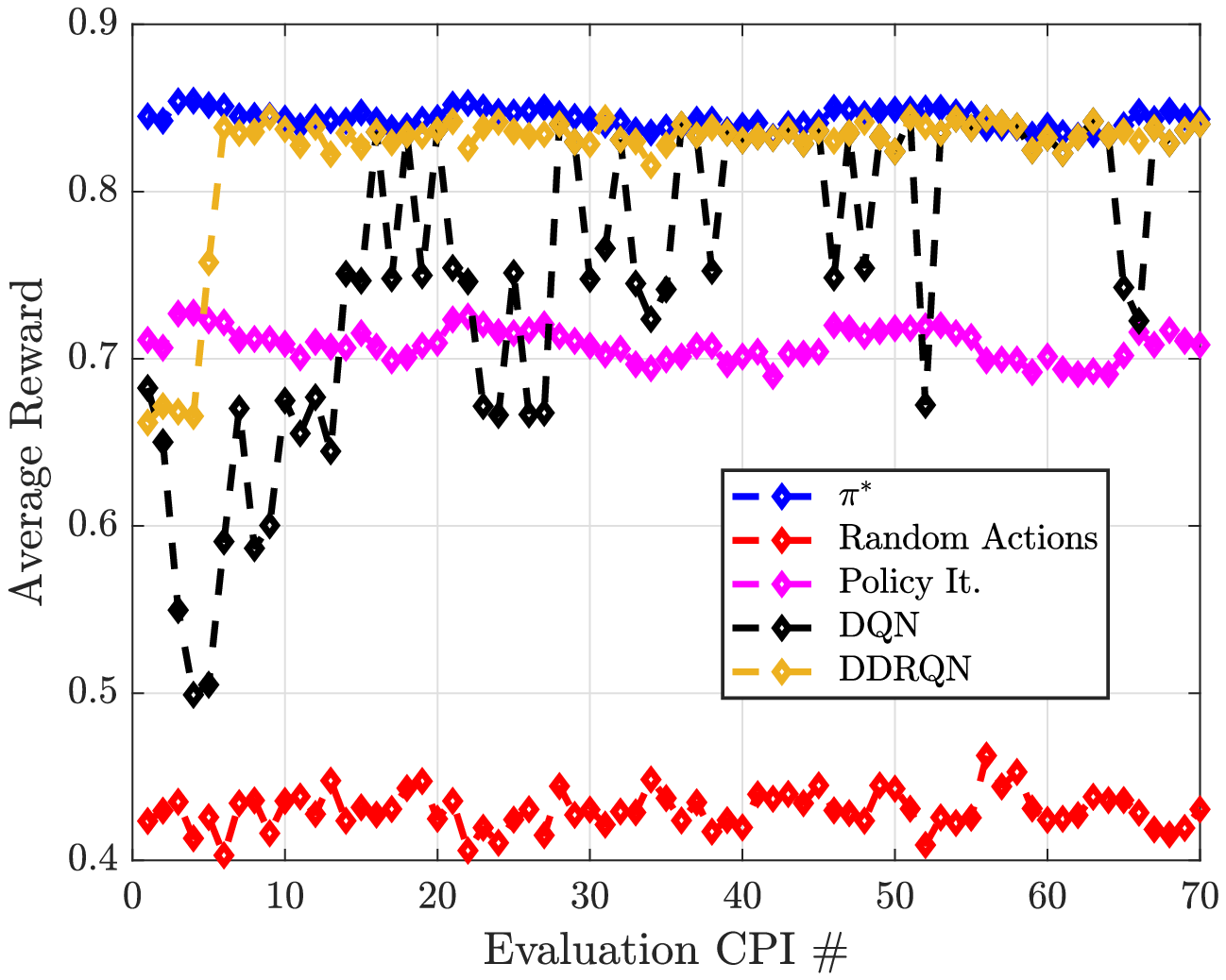}
	\includegraphics[scale=0.55]{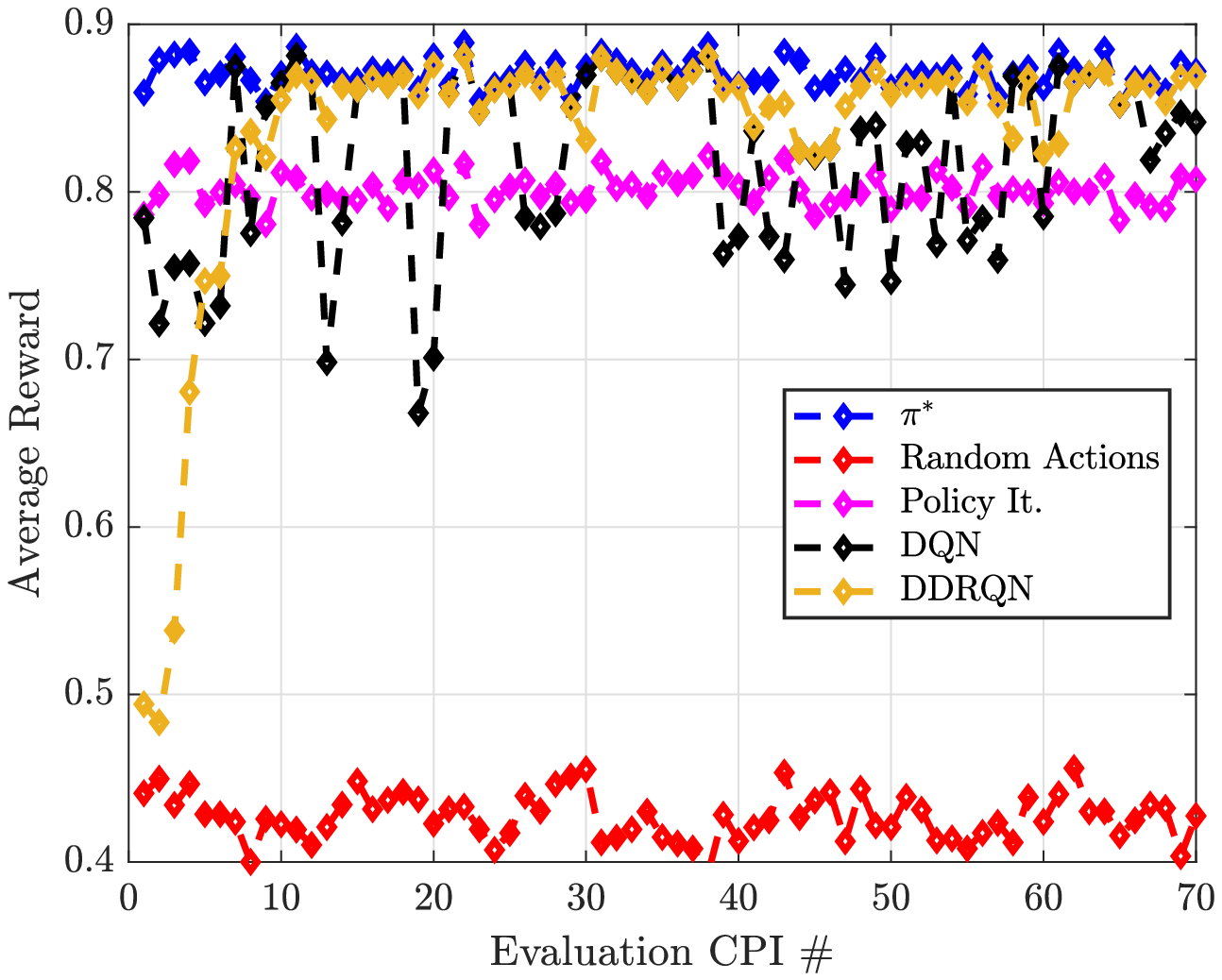}
	\caption{Average reward of each RL approach in stochastic interference generated from a Markov chain. LEFT: Interference changes state with $p = 0.4$. RIGHT: Interference changes state with probability $p=0.2$.}
	\label{fig:interact1}
\end{figure*}

\subsection{Radar Environment}
In this section, several coexistence scenarios are simulated to evaluate the performance of the Deep RL approach. The radar is evaluated in the presence of deterministic, stochastic, and recorded interference signals. For reference, the performance of the Deep RL approach is compared to the policy iteration approach and a SAA scheme, which avoids the interference seen during the last time step. The parameters used to define the radar environment and characterize the DQN and DDRQN are given in Table \ref{tab:simParam}. The radar has a pulse repetition interval (PRI) of 0.41ms, which is appropriate for long-range operation at the cost of Doppler ambiguities \cite{handbook}. Reward parameters of $\beta_{1} = 5$ and $\beta_{2} = 6$ are selected to provide a balance between interference avoidance and bandwidth utilization. A CPI of 1000 pulses is selected to observe the target's behavior over an extended time period. The target has a radar cross section (RCS) of 0.1m$^{2}$, which is characteristic of a small aircraft \cite{handbook}. The target position and velocity states are limited to $P = 50$ and $V = 10$ respectively to limit the size of the state-action space. The neural network's hyperparameters are selected via manual tuning over a variety of settings, generally following the methodology in Ch. 11 of \cite{goodfellow}.

\subsection{Radar Behavior in Stochastic Environments}
We first examine the radar's behavior in the presence of stochastically generated intermittent interference. The communications system operates in the first and second of the five sub-channels when active, or $\boldsymbol{\theta} = [1,1,0,0,0]$, and switches between `on' and `off' states according to a Markov chain model. The interference begins in the inactive state and changes state with probability $p$ at each time step independent of the history. This is an important tool for performance evaluation since the interference is non-deterministic exhibits temporal correlation dependent on parameter $p$, which is indicative of the interference seen in many wireless networks \cite{intf}.
\begin{figure*}[t]
	\centering
	\includegraphics[scale=0.50]{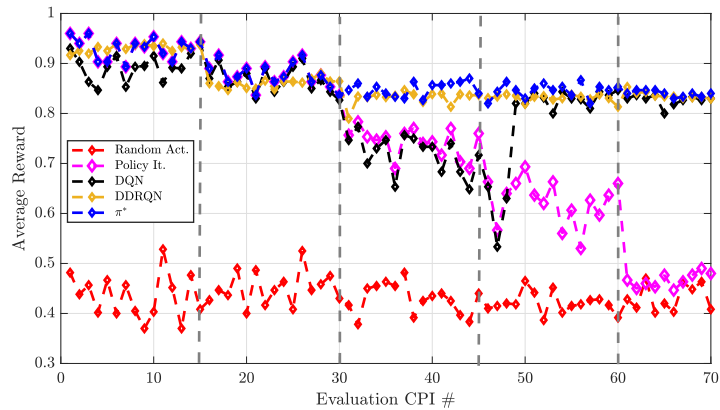}
	\caption{DQN behavior in stochastic interference generated from a non-stationary Markov model, where the probability of state changes varies over time.}
	\label{fig:stoch1}
\end{figure*}

In Figure \ref{fig:interact1}, we observe the radar's behavior in the cases of $p = 0.4$ and $p = 0.2$. In both cases we train the DQN and DDRQN models offline for 500 CPIs, and evaluate the performance over 70 additional CPIs, noting the average reward received within the CPI to examine performance and stability. For comparison, we apply the policy iteration approach to the 500 CPIs of training data and note the performance during the evaluation phase. Additionally, we analyze the performance of taking random actions, along with the $\pi^{*}$ to gain a sense of best-case and worst-case stationary performance.

In the case of $p = 0.4$, the DDRQN approach exhibits both good average performance and stability, receiving an average reward above $0.8$ upon convergence, and consistently approaches the performance of stationary $\pi^{*}$. The DQN approach achieves similar maximum performance after offline training, occasionally approaching $\pi^{*}$, but experiences a higher degree of variability, presumably due to instability in the learning process that the DDRQN is able to mitigate with the additional validation step associated with double $Q$-learning and the ability to resolve longer temporal correlations associated with recurrent $Q$-learning. It is likely that the DQN has overestimated $Q$-values associated with suboptimal actions due to maximization bias, which the DDRQN has avoided due to additional validation. Both Deep RL techniques outperform policy iteration in terms of average reward during the evaluation phase, as policy iteration is unable to establish an accurate model of $T(s,a,s^{\prime})$ and $R(s,a,s^{\prime})$ during the 500 CPIs of offline training. However, the DQN approach occasionally performs worse than the stable policy learned by policy iteration due to instability in the ongoing function approximation process.

In the case of $p = 0.2$, the interference exhibits a stronger temporal correlation between time steps, and each technique receives higher average rewards during the evaluation phase as a result. Once again, the DDRQN technique results in strong average performance and stability upon convergence, approaching stationary $\pi^{*}$ for most of the evaluation phase. Similarly, the DQN achieves good performance on average, but occasionally experiences drops in average reward due to unexpected fluctuations in the data, once again likely due to overoptimism and the inability to resolve temporal correlations. However, these drops in performance are less severe than in the case of $p = 0.4$ and very rarely result in an average reward below $0.7$ for a CPI. While the policy iteration approach receives lower average reward than either of the Deep RL techniques, the stationary policy performs relatively well compared to the case of $p = 0.4$ due to the increased stability of the interference pattern.

In dynamic coexistence scenarios, the distribution of the interference channel may change suddenly and unexpectedly. Thus, we also examine the performance of each cognitive radar approach in a non-stationary environment in Figure \ref{fig:stoch1}. In this environment, the interference is once again drawn from a two-state Markov model. However, every 15 CPIs, or 15,000 time steps, denoted by the vertical dashed lines in Figure \ref{fig:stoch1}, the probability of state change $p$ is increased as follows $p = 0.1 \rightarrow 0.2 \rightarrow 0.4 \rightarrow 0.6 \rightarrow 0.8$. 

We note from Figure \ref{fig:stoch1} that while the performance of the DDRQN model suffers slightly due to first two changes in the interference distribution, the approach eventually learns a stable policy that approaches stationary $\pi^{*}$. Thus, the DDRQN controlled radar is able to generalize smoothly to the more drastic interference changes later in the evaluation interval. The DQN approach similarly learns an effective policy by the end of the evaluation interval, but experiences performance degradation for approximately 50 CPIs before converging to stable behavior. The policy iteration approach experiences increasingly drastic drops in performance with each change in the interference distribution, exhibiting close to worst-case behavior once the interference parameter $p = 0.8$. 
\begin{table*}[t]
	\footnotesize
	\centering
	\caption{Comparison of cognitive radar approaches in simulation. The Deep RL (DQN) approach is compared to MDP Policy Iteration and Sense and Avoid (SAA) methods.}
	\label{tab:t1}
	\begin{tabular}{|l|l|l|l|l|l|}
		\hline
		Interference (CR approach) & Av. SINR (dB) & Av. BW (MHz) & \% Coll. & \% Missed Opp. & \% Adapt. \\ \thickhline
		FH Sweep (SAA) & -5.19 & 64.0 & 60.0\% & 40.0\% & 100.0\% \\ \hline
		FH Sweep (Policy It.) & 22.04 & 80.0 & 0.0\% & 0.0\% & 100.0\% \\ \hline
		FH Sweep (DQN) & 22.04 & 80.0 & 0.0\% & 0.0\% & 100.0\% \\ \hline
		FH Sweep (DDRQN) & 22.04 & 80.0 & 0.0\% & 0.0\% & 100.0\% \\ \thickhline
		M.C. (SAA) & 11.42 & 80.64 & 22.58\% & 22.58\% & 21.2\% \\ \hline
		M.C. (Policy It.) & 19.60 & 55.08 & 3.8\% & 65.8\% & 20.6\% \\ \hline
		M.C. (DQN) & 22.01 & 60.0 & 0.0\% & 48.9\% & 0.0\% \\ \hline
		M.C. (DDRQN) & 21.85 & 63.04 & 0.25\% & 42.6\% & 11.4\% \\ \thickhline
		2450MHz (SAA) & 2.79 & 66.94  & 41.4\% & 35.9\% & 73.8\% \\ \hline
		2450MHz (Policy It.) & 7.54 & 58.18 & 30.2\% & 51.0\% & 59.9\% \\ \hline
		2450MHz (DQN) & 14.95 & 35.78 & 15.9\% & 74.80\% & 28.90\% \\ \hline
		2450MHz (DDRQN) & 15.65 & 39.4 & 12.6\% & 72.20\% & 31.30\% \\ \hline
	\end{tabular}
\end{table*}

In this section we have demonstrated the ability of a Deep RL controlled cognitive radar to coexist with a both stationary and non-stationary stochastic interference sources in terms of convergence in the average received reward. Given a fixed offline training period of 500 CPIs, the proposed Deep RL approaches perform favorably compared to policy iteration as an explicit expression of $T(s,a,s^{\prime})$ and $R(s,a,s^{\prime})$ is not required. To further examine the utility of the proposed Deep RL approaches, we observe important radar performance metrics in the next section.

\subsection{Comparison of Radar Performance Metrics}
In this section, we compare radar performance metrics achieved using the Deep RL approach to policy iteration and a basic Sense-and-Avoid (SAA) scheme. The SAA approach selects the largest contiguous group of sub-channels in the previously observed environmental state and does not learn from observations beyond time $t-1$. However, SAA is computationally efficient and may be more effective than a RL approach when environmental patterns are difficult to predict or very quick responses to changing environmental dynamics are necessary. Additionally, SAA may be effective when immediate performance guarantees are necessary, as no exploration phase is required.

We consider the radar's performance in deterministic and stochastic environments in addition to spectrum data recorded from the 2450MHz band. The radar's performance is evaluated in terms of average received SINR, collisions, missed opportunites, and number of waveform adaptations. We compute SINR using the following expression
\begin{equation}
\operatorname{SINR} =\frac{P_{S}}{P_{I}+P_{N}}=\frac{P_{T} G_{T} G_{R} \lambda^{2} \sigma}{((4 \pi)^{3} R^{4} k T_{0} L) + \frac{N_{c}}{N}P_{I}},
\end{equation}
\noindent where $P_{T}$ is the radar's transmission power, $G_{T}$ and $G_{R}$ are the transmit and receiver antenna gains, $\lambda$ is the wavelength, $\sigma$ is the target radar cross section (RCS), $R$ is the target range, $k$ is the Boltzmann constant, $T_{0}$ is the noise temperature, $L$ is a general loss factor, and we assume that the interference power at the radar $P_{I}$ is known in advance from an initial passive sensing period where a blind estimation technique such as energy detection is used. This assumption is reasonable if the interferer's transmit power is held constant over the observation interval.
\begin{definition}
	A waveform adaptation is said to occur whenever $\mathbf{a}_{t} \neq \mathbf{a}_{t-1}$. This corresponds to differing LFM waveform parameters between adjacent pulses.
\end{definition}
Although the proposed approach does not attempt to optimize the rate of adaptation, it is an important parameter as significant waveform adaptation within a CPI can lead to incoherent received data, complicating target detection and tracking.

The first interference type used for comparison is a frequency-hopping sweep, which occupies one sub-band at each time step and moves across the channel in a repeating pattern as follows: $\boldsymbol{\theta} = [1,0,0,0,0] \rightarrow [0,1,0,0,0] \rightarrow [0,0,1,0,0] \rightarrow [0,0,0,1,0] \rightarrow [0,0,0,0,1]$. Secondly, we use a Markov chain interference model that occupies $\boldsymbol{\theta} = [1,1,0,0,0]$ when active, as in the previous section. This interference source begins in the `off' state and switches state with probability $p = 0.4$. Test and training data are drawn independently from the Markov chain. Lastly, we use a recorded sample of RFI from a 100MHz channel centered around 2450MHz, which contains unlicensed WiFI and Bluetooth signals. This recorded data is split up into training and evaluation segments, which originate from the same data collection. The RL agents use the a common reward function, given by (\ref{eq:rwd}).

In Table \ref{tab:t1}, we see performance metrics for each interference type. We observe that for the case of the \textit{deterministic frequency hopping sweep}, each of the RL techniques is able to learn the one-step $T(s,a,s^{\prime})$ and $R(s,a,s^{\prime})$ dynamics explicitly, and optimal performance in terms of missed opportunities and collisions is achieved. The SAA approach however, performs very poorly in this case as $\boldsymbol{\theta}_{t} \neq \boldsymbol{\theta}_{t+1}, \; \; \forall \; t,$ which violates SAA's assumption that $\boldsymbol{\theta}_{t} = \boldsymbol{\theta}_{t+1}$. Thus, we see that in cases where $T(s,a,s^{\prime})$ is deterministic, all RL approaches are effective given enough training time. However, Deep RL techniques may be beneficial over a policy iteration approach if the training time is limited, as an explicit transition and reward model is not required. Further, if the transition probabilities are deterministic over many states, the DDRQN may be able to better resolve these dependencies due to the recurrent LSTM architecture.

In the presence of the \textit{stochastic Markov chain} interference with $p = 0.4$, the performance of the SAA approach is improved compared to the FH Sweep. This is because, on average, the interference is stationary for 1.5 time steps and often remains stationary for much longer. However, the policy iteration technique performs much better in terms of interference mitigation, as $R(s,a,s)$ gives the minimum reward whenever $N_{c} \geq 0$. The DQN approach learns that a high reward can be achieved by avoiding the sub-bands containing interference completely, and avoids all collisions with no waveform adaptations. The DDQN also avoids nearly all collisions while utilizing slightly more bandwidth, resulting in fewer missed opportunities at only a slightly lower average SINR than DQN.

The \textit{2450MHz recorded} interference contains multiple signals transmitting during the data collect and is reflective of a crowded spectral environment. Here, the interference is largely non-coherent between time steps and the SAA approach performs poorly in terms of both SINR and bandwidth utilization. The policy iteration approach slightly reduces the rate of collisions, but results in a lower bandwidth utilization than SAA. Further, the Deep RL approaches are able to significantly reduce the number of collisions compared to policy iteration, resulting in significant SINR improvement compared to the other techniques, with the DDRQN achieving the lowest rate of collision. However, since the environment's transition dynamics are not explicit, the rate of missed opportunity is significant for both techniques.

We have thus observed that reward-driven control of a cognitive radar system improves radar performance metrics in many coexistence scenarios and is generally favorable compared to a reactive approach. In particular, Deep RL is a viable tool for learning the environment's dynamics without explicit representations of $T(s,a,s^{\prime})$ and $R(s,a,s^{\prime})$. This allows for improved generalization performance compared to the tabular policy iteration approach. Further, the use of double learning and a RNN architecture affords a cognitive radar stability in the learning process and the ability to resolve dependencies over many time-steps.

\section{Hardware Implementation and Experimental Results}
In this section we present a novel implementation of Deep RL cognitive radar control on a Software Defined Radar (SDRadar) prototype. The radar is developed using a Universal Software Radio Peripheral (USRP) X310 model, which is selected for its low cost and flexibility. The USRP's UBX-160 RF daughterboards provide sufficient performance for making high-quality radar measurements over the 100MHz instantaneous bandwidth of interest, which is used to test the coexistence scenario that was simulated in the previous section. The SDRadar setup utilizes a host PC which provides the system with user control and performs the training of the Deep RL agent as well as other processing operations.

The fast spectrum sensing (FSS) algorithm is used to process spectrum data in real time. The FSS method quickly determines the vacancy of sub-bands by emulating the rapid data processing of the human thalamus and is described in detail in \cite{noncoop}. Once FSS is performed, the spectrum data is then split into five equally spaced sub-bands to form the binary interference state vector $\boldsymbol{\theta}$ at each time step. To train the DQN model, we use only a passive sensing period or \textit{offline training}, during which the radar observes the RF environment and simulates taking random actions to estimate \textit{Q}-values for each action. This is done to save computational resources so that real-time waveform selection can be performed. The \textit{Q}-values are obtained by using the estimated SINR value and bandwidth utilized during each action to calculate a reward. Based on the \textit{Q}-values calculated during the sensing period, the radar then compiles a table that determines which action to take for any scenario. During real-time operation, we are limited by the timing constraint
\begin{equation}
T_{WS} <  \frac{T_{Int}}{2}.
\end{equation}
\noindent Where $T_{WS}$ is the time required for waveform selection and $T_{Int}$ is the interference time slot. Since we are concerned with adapting our waveform quickly, picking an action directly from a Look-Up Table (LUT) generated from a frozen DQN allows us a quick waveform selection time $T_{WS}$. However, online $Q(s,a)$ updates are not considered here. Future work will focus on reducing the computational resources of signal processing tasks so online training can be performed.
\begin{table}[t]
	\footnotesize
	\centering
	\caption{Average SINR values for 0.41s of SDRadar data using SAA, policy iteration, and DQN approaches. Communications signals are used as RFI. }
	\label{tab:tab2}
	\begin{tabular}{lllll}
		\multicolumn{5}{c}{SINR (dB)} \\ \hline
		\multicolumn{2}{|l|}{RFI Type} & \multicolumn{1}{l|}{SAA} & \multicolumn{1}{l|}{Policy Iteration} & \multicolumn{1}{l|}{DQN} \\ \hline
		\multicolumn{2}{|l|}{LTE FDD} & \multicolumn{1}{l|}{23.08} & \multicolumn{1}{l|}{24.42} & \multicolumn{1}{l|}{29.72} \\ \hline
		\multicolumn{2}{|l|}{LTE TDD} & \multicolumn{1}{l|}{12.60} & \multicolumn{1}{l|}{18.023} & \multicolumn{1}{l|}{28.56} \\ \hline
		\multicolumn{2}{|l|}{GSM} & \multicolumn{1}{l|}{26.95} & \multicolumn{1}{l|}{28.893} & \multicolumn{1}{l|}{30.02} \\ \hline
		\multicolumn{2}{|l|}{CDMA} & \multicolumn{1}{l|}{14.89} & \multicolumn{1}{l|}{29.56} & \multicolumn{1}{l|}{29.78} \\ \hline
	\end{tabular}
\end{table}

Once the offline training phase is complete, the radar then acts by utilizing the best action for each scenario, which is done by observing the interference state as well as some number of previous states and using the LUT to choose the best action. Since a LUT is established before radar transmits begin, the radar is able to save computational resources for other tasks to complete real-time processing. Since this implementation only utilizes offline training, if the environment changes drastically the network will need to be retrained. However, our results show that the DQN approach does yield better performance in terms of RFI avoidance than a hardware implementation of the MDP policy iteration algorithm and a simple SAA method.
\begin{figure}[t]
	\centering
	\includegraphics[scale=0.30]{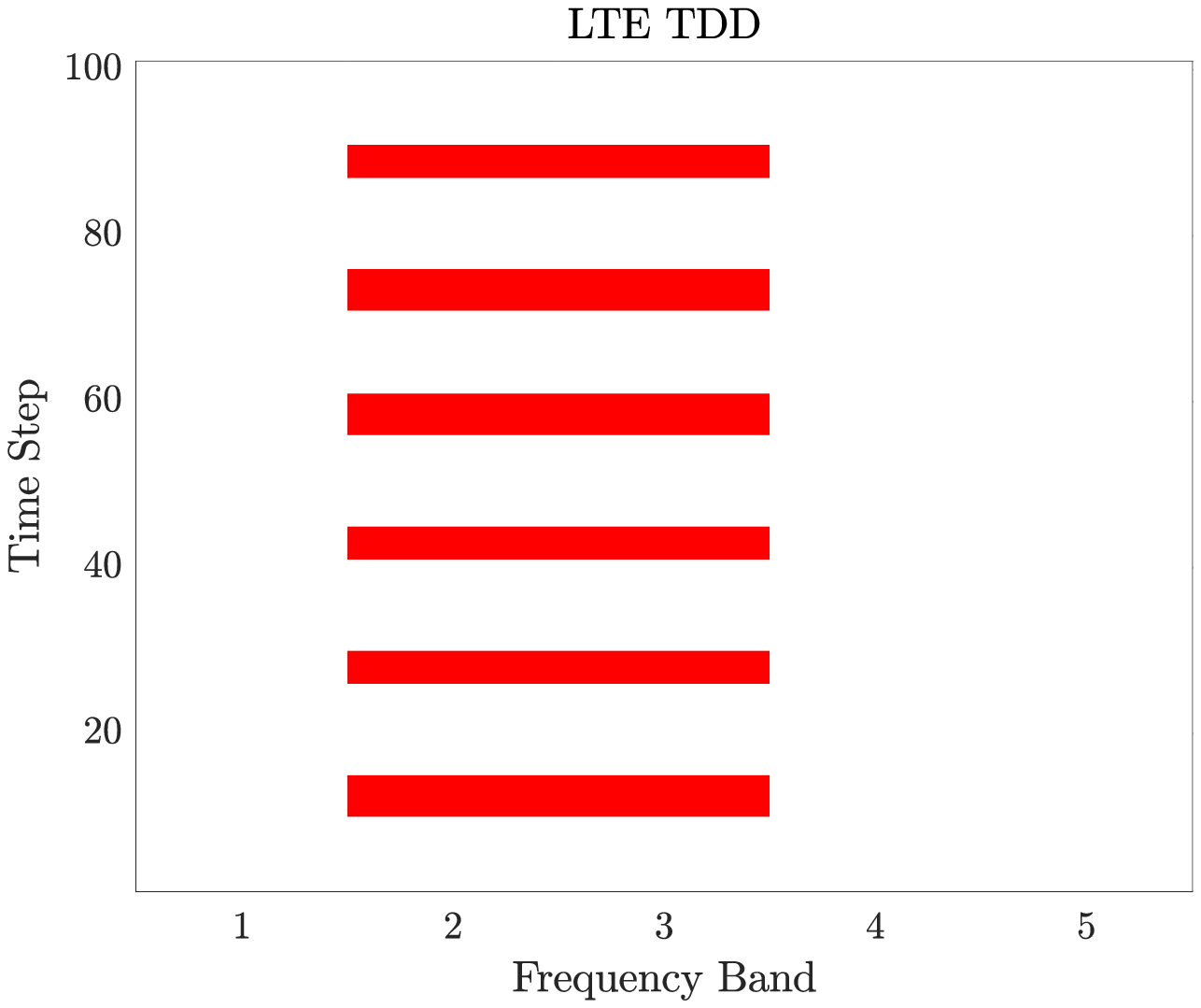}
	\includegraphics[scale=0.30]{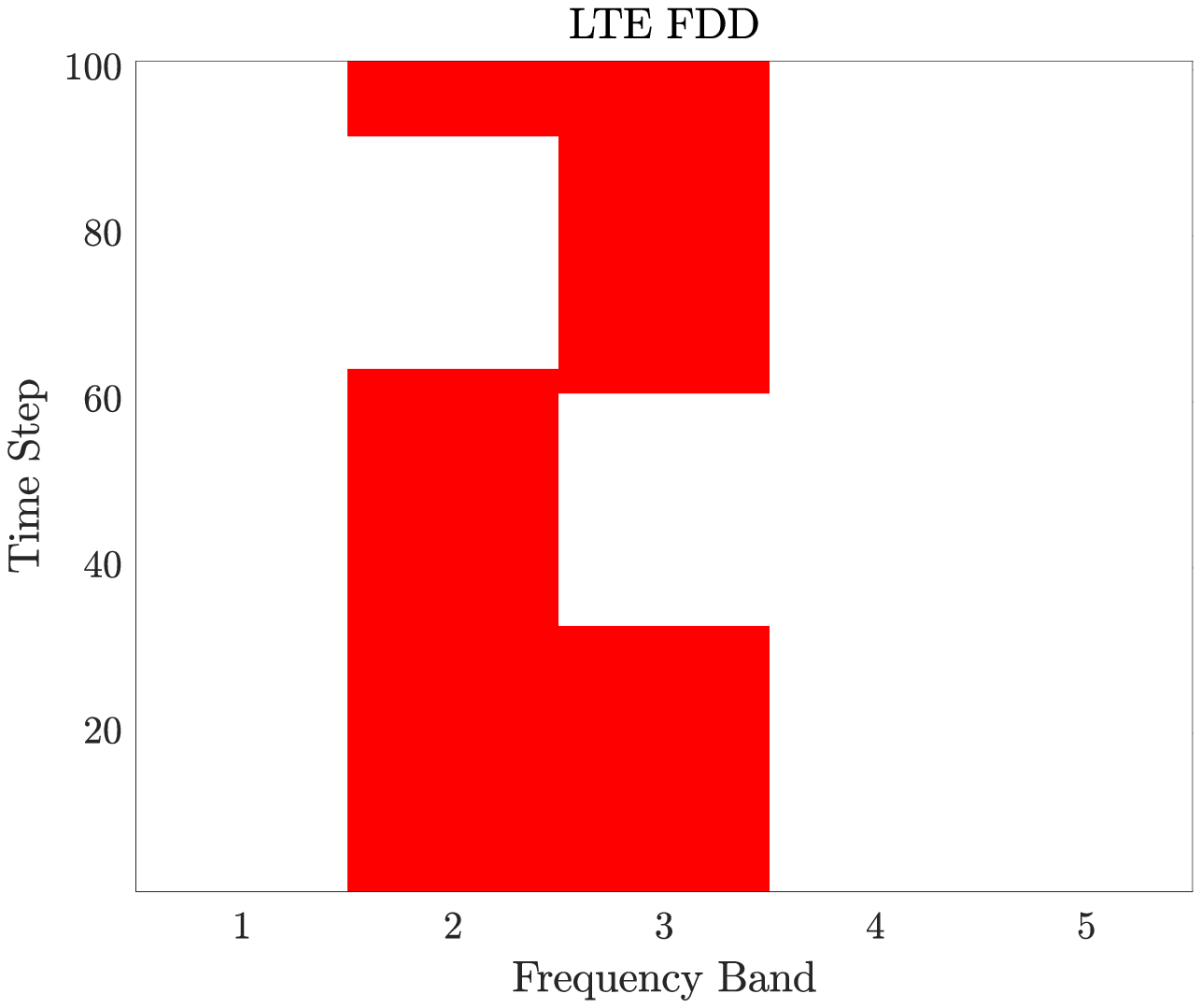}
	\includegraphics[scale=0.30]{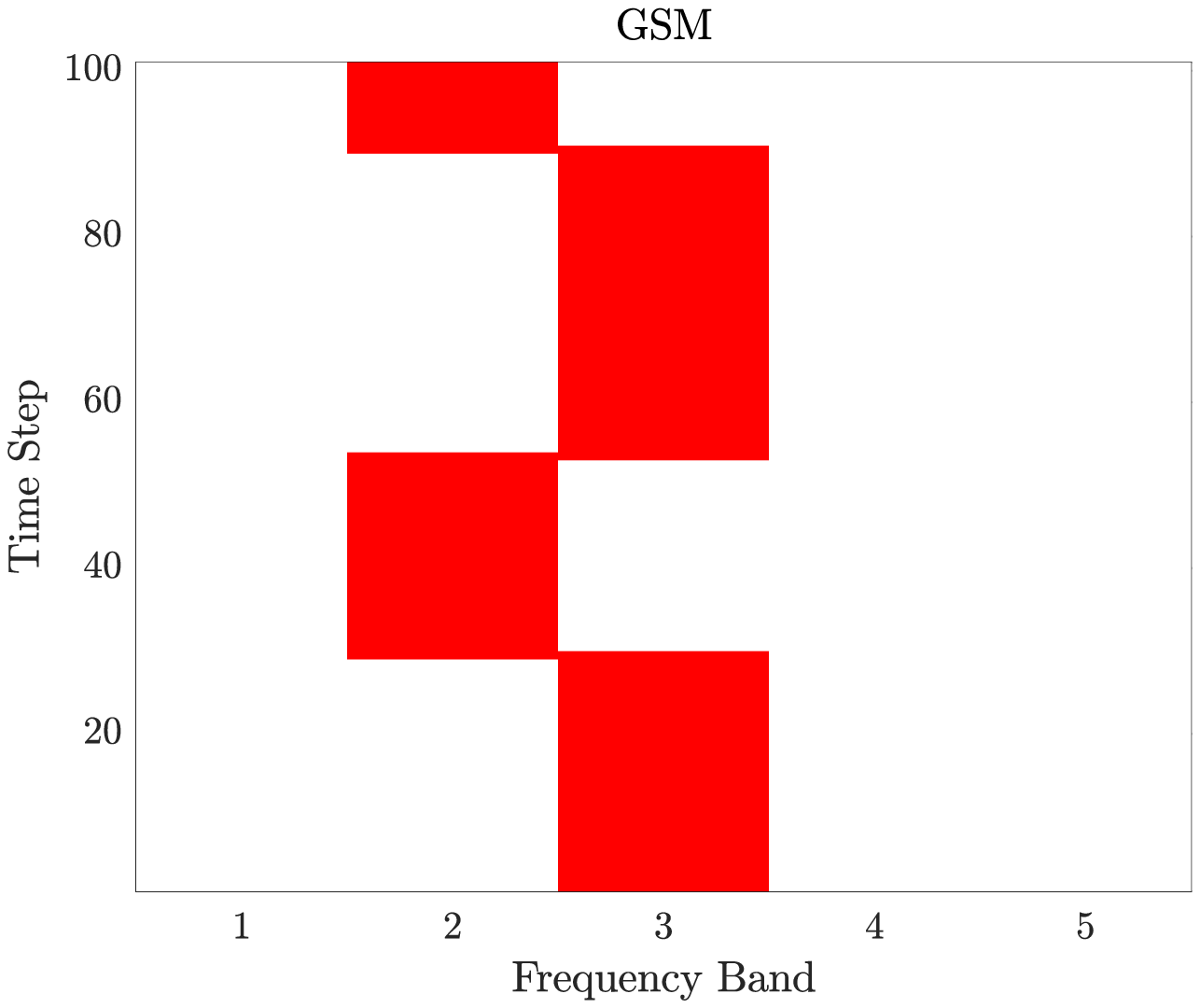}
	\includegraphics[scale=0.30]{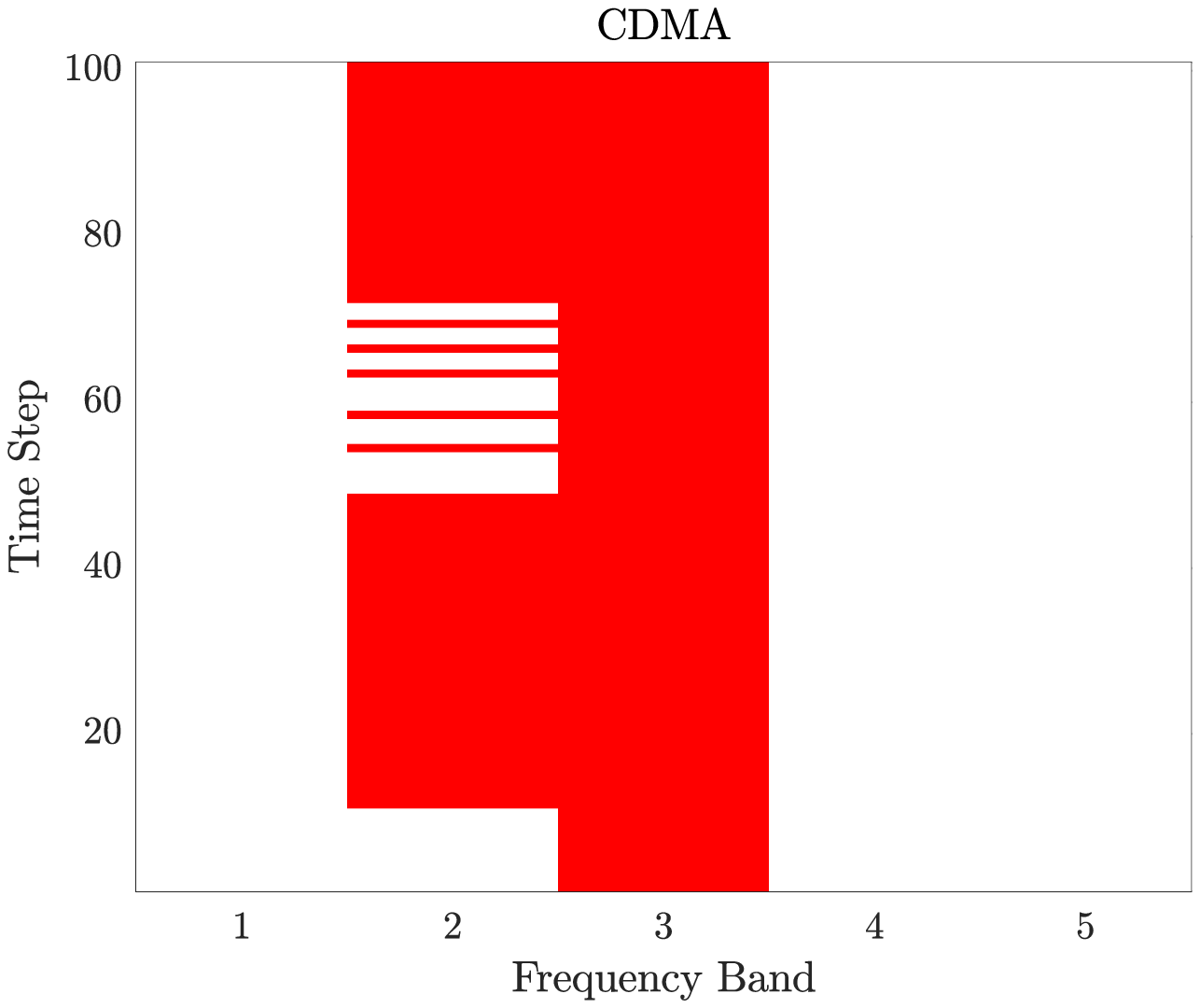}
	\caption{Binary representations of recorded communication signals used to test the radar}
	\label{fig:rfi}
\end{figure}

To introduce interference to the environment we use a Vector Signal Transceiver (VST) as an arbitrary waveform generator to play back interference recorded in the physical world or generated synthetically. The interfering signal is added to the radar's LFM chirp waveform with an RF combiner. The SDRadar is connected to a host computer via a 4x PCIe data transfer cable, which provides low latency communication between the PC and SDR. Processing in this system occurs primarily on the host PC, but also uses the FPGA of the USRP for some signal processing tasks.
\begin{figure*}[t]
	\centering
	\includegraphics[scale=0.42]{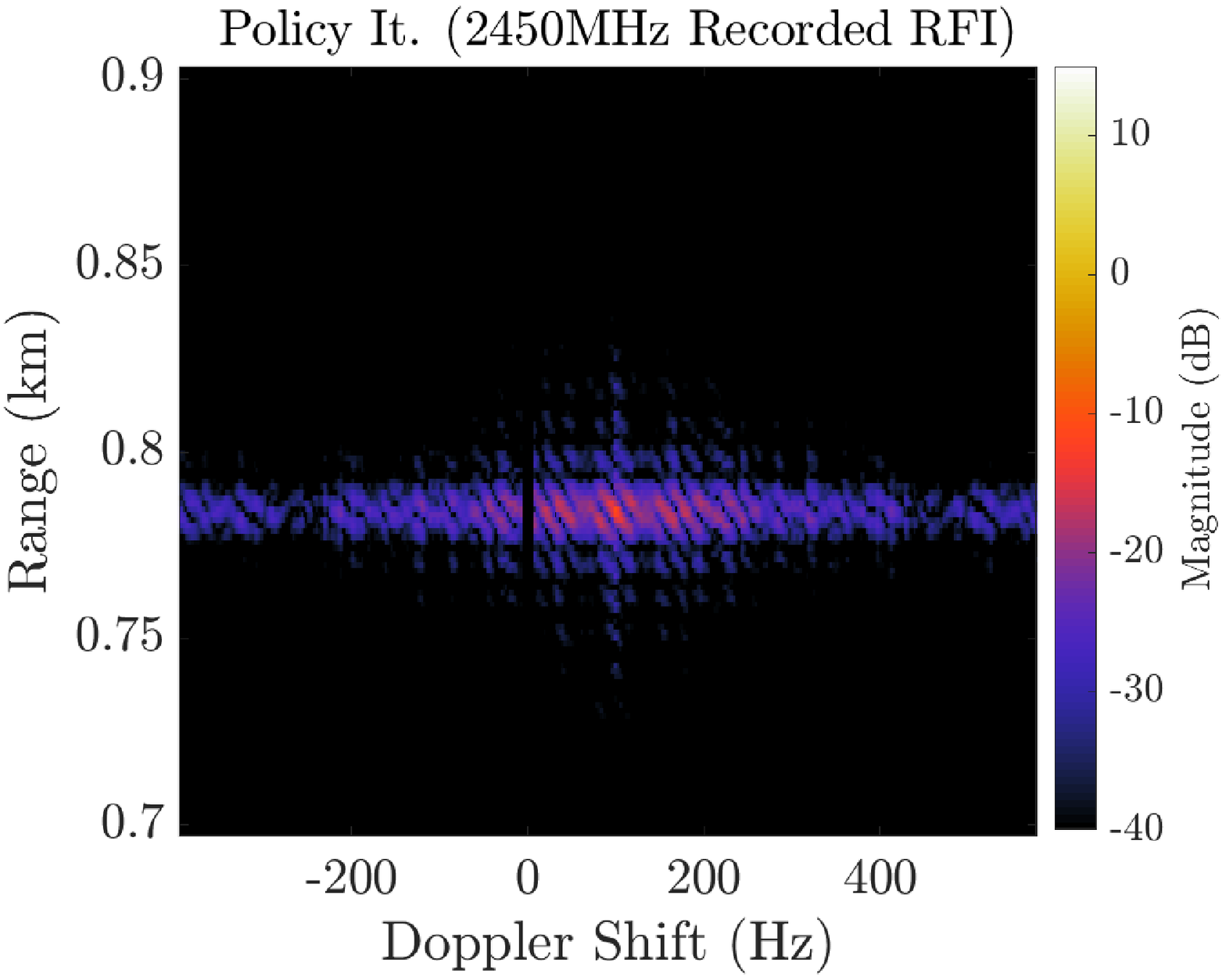}
	\includegraphics[scale=0.42]{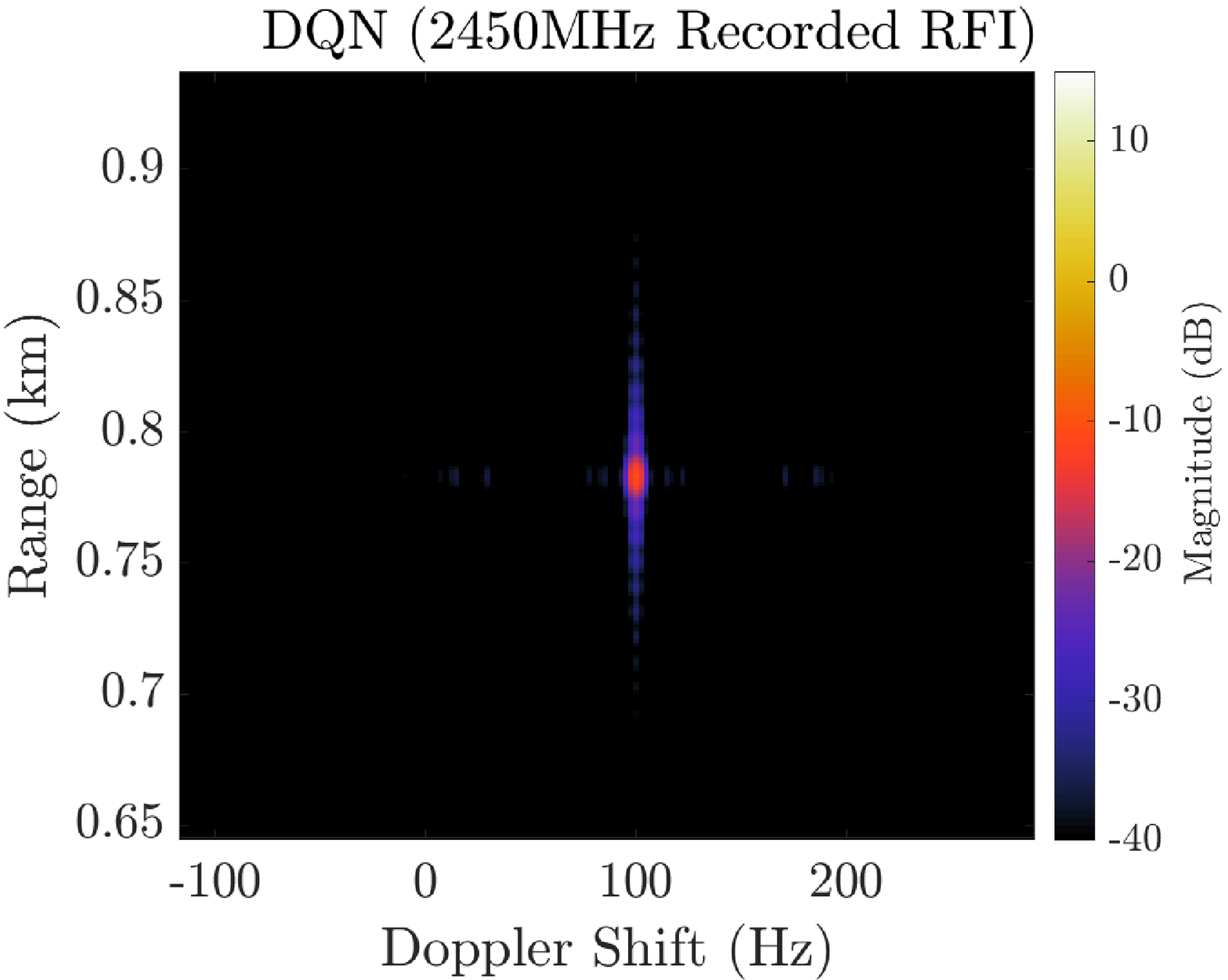}
	\caption{LEFT: Range Doppler response of policy iteration approach avoiding recorded interference from the 2450MHz band. The high Doppler sidelobes are a result of switching the chirp waveforms center frequency within the CPI. RIGHT: Range-Doppler response of DQN-enabled radar avoiding recorded RFI. The DQN approach varies its waveform less often, resulting in lower Doppler sidelobes.}
	\label{fig:radar}
\end{figure*}

\subsection{Experimental Results}
Here, we are once again interested in utilizing as much of the available spectrum as possible without causing mutual interference. However, our main objective is to improve radar detection and tracking performance through this approach, so we are also interested in the range-Doppler response of coherently processed data. We thus examine the received SINR and bandwidth statistics to analyze radar spectrum utilization. Addtionally, we utilize range-Doppler plots and the Constant False Alarm Rate (CFAR) algorithm to examine the radar's detection performance. 

Feedback from the radar processing dictates how the radar's rewards should be best tailored to fit a particular environment, while the rewards themselves give an indication of how well the radar is performing on the task it is assigned. While designing a reward function based solely on the radar processing performance would be useful, getting immediate feedback regarding the radar performance of individual actions can be difficult, as coherent processing must be performed. Therefore, we utilize the reward function to mitigate missed opportunities and collisions on a pulse-to-pulse basis.
\begin{figure*}[t]
	\centering
	\includegraphics[scale=0.5]{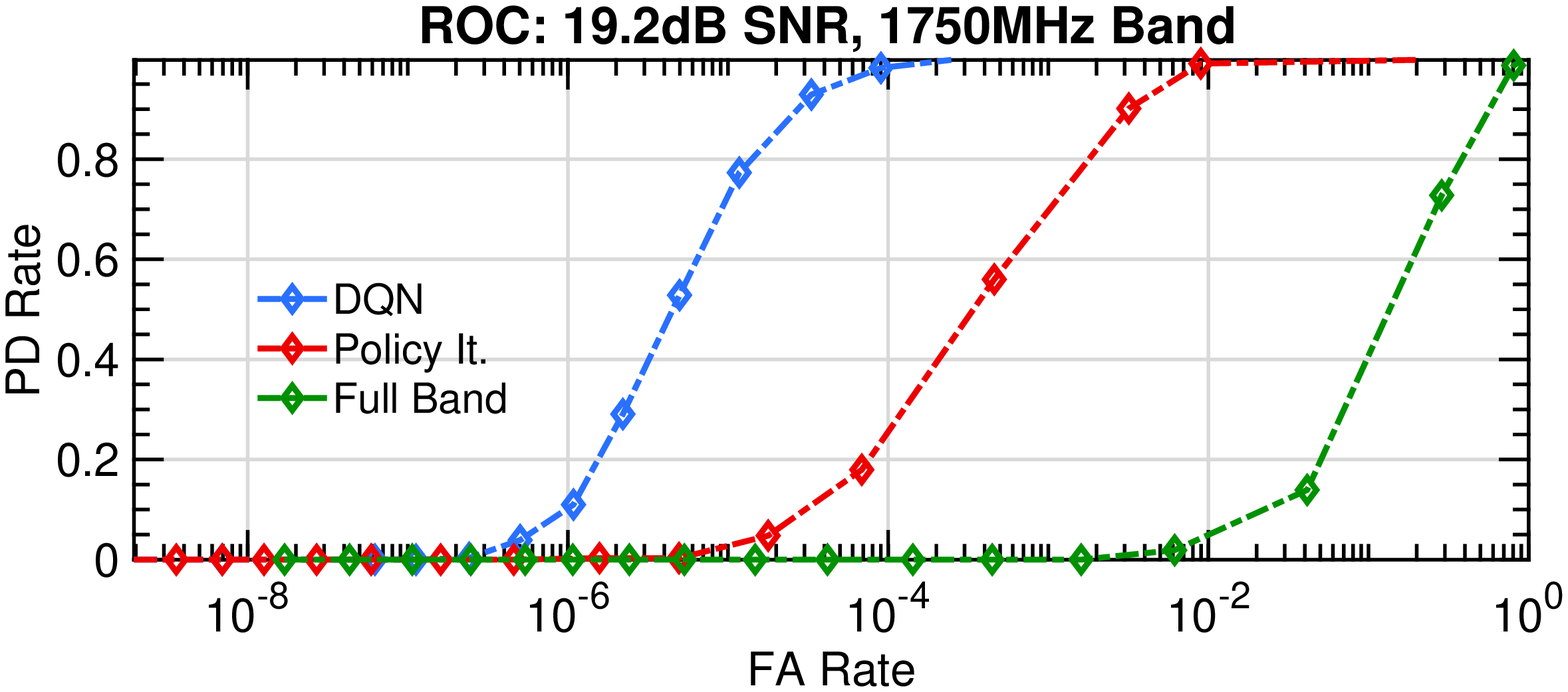}
	\includegraphics[scale=0.5]{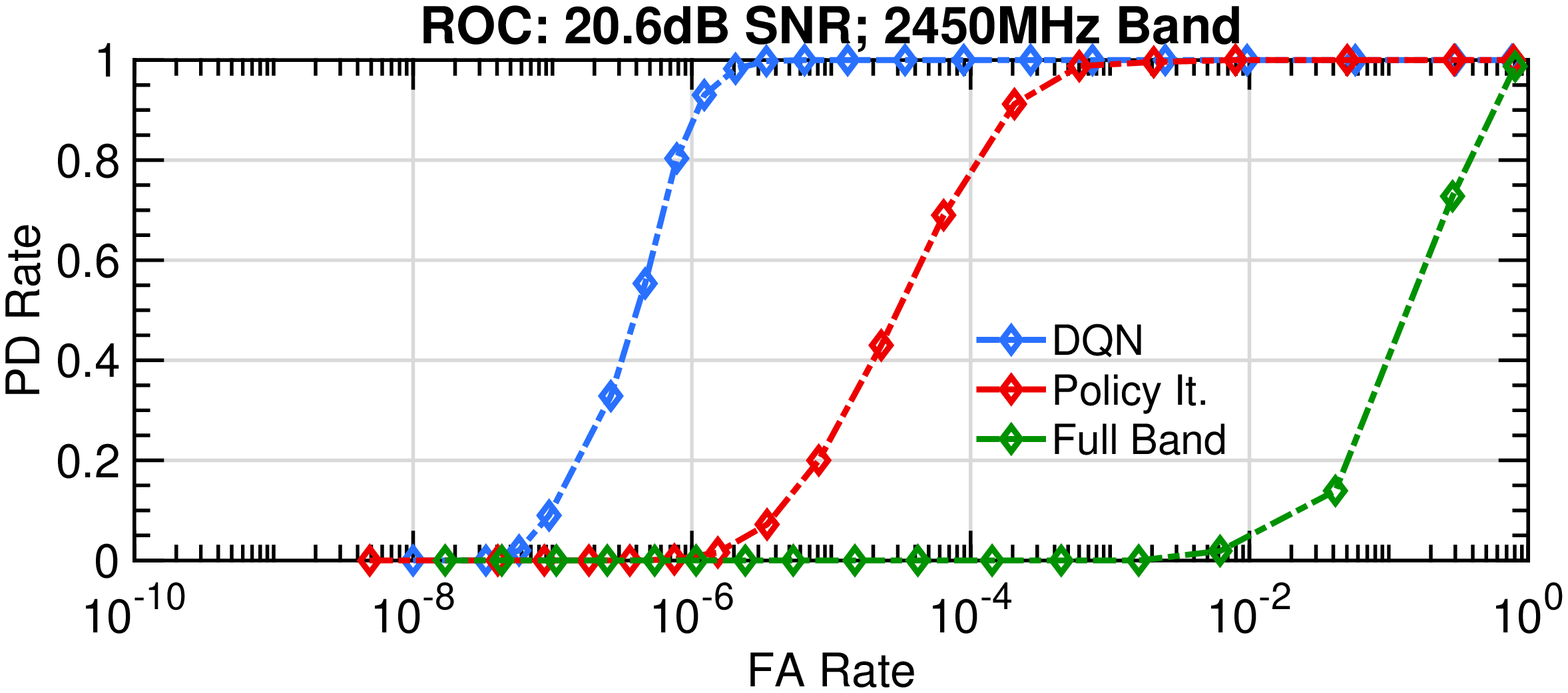}
	\caption{ROC curves for CFAR detection using 500 CPIs of radar data for each approach when recorded RFI is present. The DQN (blue) demonstrates increased positive detections for a given false alarm rate compared to the policy iteration approach (red) and traditional radar (green). }
	\label{fig:1750}
\end{figure*}

This method was evaluated by direct comparison with a hardware implementation of the MDP policy iteration approach as well as a SAA approach with no predictive capabilities. RFI avoidance for each method is evaluated after the radar returns from each method is coherently processed and matched filtered in a post-processing operation. Here the Pulse Repetition Interval (PRI) is $409.6 \; \mu s$ and a CPI consists of 1000 pulses $(0.41s)$. The duration of the LFM chirp is 20 $\mu$s. The RFI considered for this comparison are binary representations of recorded time-varying communications signals, which include LTE Frequency Division Duplexing (FDD), LTE Time Division Duplexing (TDD), GSM, and CDMA schemes, samples of which can be seen in Figure \ref{fig:rfi}.

In Table \ref{tab:tab2}, we see the average SINR values after coherent processing and matched filtering operations for the SAA approach along with the MDP policy iteration and DQN cognitive engines. For the FDD and GSM cases, the reactive mode is able to avoid most of the RFI, but still experiences some collisions as it cannot predict transitions. Both policy iteration and DQN provide small SINR benefits for these cases as the need for predictive power in these scenarios is present but limited. 

The SINR improvement from the DQN is most noticeable for the LTE TDD data, where roughly a 10dB improvement is seen from the policy iteration approach, which is in turn a roughly 8 dB improvement over the reactive method. The DQN provides such a large improvement over policy iteration due to the benefit of additional memory. Since the interference is following a predictable pattern, the DQN is able to capture this by looking at the past three RFI states as opposed to policy iteration, which can only look at a single state when making decisions.

Another potential advantage of the DQN approach in a hardware implemented system is that the radar tends to select the same action many times if it reliably provides a high \textit{Q}-value. If the radar is trained to convergence, we tend to see stable average \textit{Q}-values. This generally results in less adaptation of the transmit waveform, which may help mitigate sidelobes in the Doppler domain. In Figure \ref{fig:radar}, the range-Doppler response of policy iteration and DQN-enabled radars avoiding 100MHz of recorded RFI from the 2450MHz center frequency band is seen. The target is much more visible when the DQN policy is utilized compared to the policy iteration approach, as the processed data being more coherent. While distortion effects from pulse agility may be corrected by signal processing techniques, such as the Richardson-Lucy deconvolution algorithm or non-identical multiple pulse compression (NIMPC) \cite{richardson}, \cite{nimpc} the DQN approach may mitigate the need for computationally expensive post-processing.

To quantitatively evaluate the performance of the Deep RL in the hardware implementation, we also examined the radar detection performance and compare with the policy iteration algorithm and a traditional radar scheme, which always uses the full band of interest. We compared detection performance by using the CFAR algorithm and generating Receiver Operating Characteristic (ROC) curves, which consider the Positive Detection, $PD$, and False Alarm, $FA$, rates for various target detection thresholds. Each point on the ROC corresponds to 500 CPIs of received radar data, each 0.41s in duration. For each point, we use CFAR to calculate a theoretical false alarm rate which determines the detection threshold. The rates of $PD$ and $FA$ are measured as follows:
\begin{align}
\operatorname{PD \; Rate} &= 1 - \frac{1}{N} \sum_{j = 1}^{N} \operatorname{MD}_{j}\\
\operatorname{FA \; Rate} &= \frac{1}{N \times N_{P}} \sum_{j = 1}^{N} \operatorname{FA}_{i},
\end{align}
\noindent where $N$ is the number of CPIs examined, $\operatorname{MD_{j}}$ is the number of detections missed, assuming we know that a target is present in the environment, and $N_{P}$ is the number of points in the 2D range-Doppler map. The results of the CFAR analysis for the radar operating in 100MHz of RFI recorded from bands centered at 1750MHz and 2450MHz in a campus environment can be seen in Figure \ref{fig:1750}.

We thus observe that the using RL agents for radar control results in a significant shift of the performance curve to the left in ROC space compared to traditional radar operation, demonstrating a decreased amount of false alarms for a positive detection rate. Additionally, the DQN approach results in a significant performance increase from the policy iteration approach. This is due to the increased SINR in the coherently processed data by using the DQN to avoid RFI and lower Doppler sidelobes in the range-Doppler map due to the smoother action pattern the Deep RL agent takes in each scenario.

Here we have proposed the DQN agent as a viable cognitive engine for avoidance of time-varying RFI and increased target detection performance in congested spectrum. Although the radar must be trained prior to operation, once the training is complete the radar is able to quickly act on the best actions without significant processing demands, providing efficient operation in a real-time system. Additionally, the radar is able to predict and avoid time-varying RFI more accurately than the previously proposed policy iteration method, resulting in higher average SINR for range-Doppler processed data and increased detection performance when the CFAR algorithm is used, which can be seen in the ROC space through a decreased number of false alarms for a given amount of true detections. Future work on this implementation will focus on reducing the processing done by the host PC such that online training can occur during radar operation. 

\section{Conclusion and Future Directions}
This work has presented and demonstrated an effective Deep RL framework for a spectrum sharing cognitive radar system. The proposed approach uses non-linear function approximation via deep neural networks to estimate the value of state-action pairs directly from the radar's experiences. The proposed learning scheme utilizes Deep $Q$-Learning to stabilize the function approximation procedure through the use of experience replay and a slow-updating target network. Further, it has been demonstrated that an extension to DQL that utilizes double learning in tandem with a RNN architecture is able to achieve very good average performance and stability in the presence of both stationary and non-stationary stochastic interference.

The cognitive radar system is able to find a solution that improves detection performance while acting in a congested spectral environment by developing behavior to maximize a reward function. Thus, design of a reward function that accurately motivates desired radar behavior is a major concern when utilizing this method. We have shown through simulation and experimental results that a reward function that primarily learns to avoid collisions with RFI, and secondarily learns to utilize as much contiguous bandwidth as possible given that no collisions occur, is an effective learning strategy to improve radar performance metrics.

The Deep RL approach has shown considerable radar performance improvement over the previously proposed MDP policy iteration approach, which is seen both in simulation and experimental results from a hardware prototype. The largest benefit is the reduced dimensionality required to develop a solution as compared to policy iteration, as the radar does not need to learn the explicit transition and reward models of the environment. This allows the radar to learn more efficiently in cases where the environment is difficult to characterize. Further, the Deep RL scheme is able to scale to larger state-action spaces more effectively, allowing for more CSI to drive the radar's decisions in complex scenarios. 

Another benefit of the Deep RL approach is that function approximation is used, allowing the radar to generalize more effectively than the policy iteration approach, which suffers in performance when the environment's one-step dynamics change. Utilizing function approximation updates the value of many state-action pairs based on a single experience, allowing for better generalization performance when previously unseen states appear.  

While Deep RL is an effective and practical method to enable spectrum sharing radar, there are several associated challenges. The first is that deep neural networks often require significant exploration to learn the large number of weights necessary for sufficient performance. Additionally, variation of the chirp waveform's bandwidth and center frequency regularly within a CPI can lead to significant distortion in the Doppler domain \cite{richardson}. A hierarchical learning scheme which learns from range-Doppler processed data in addition to feedback from radar pulses may be an effective way to mitigate this effect and could be investigated in future research. Hierarchical learning for cognitive radar has been briefly explored in \cite{metacog}, but the empirical success of hierarchical RL algorithms \cite{barto} necessitates further study in the context of dynamic spectrum sharing. Additionally, this work utilized an extended offline training period where the radar explores the effect of taking random actions. However, in applications where minimum performance guarantees must be met during training, an \textit{on-policy} approach, such as SARSA \cite{sutton}, may be warranted and could be a subject of future study.

Deep RL presents a promising path forward for control of cognitive wireless systems that must operate in increasingly congested spectrum sharing scenarios. While there is still significant work to be undertaken to develop robust algorithms suitable for a generic RF environment, this approach presents a viable solution which can be combined with domain knowledge as a step towards more general, autonomous, and intelligent cognitive radar systems.

\bibliography{jbib}{}
\bibliographystyle{IEEEtran}

\end{document}